\documentclass[journal=ancac3,manuscript=article,chaptertitle=true,articletitle=true]{achemso}
\setkeys{acs}{articletitle=true}
\usepackage[utf8]{inputenc}
\usepackage[version=3]{mhchem} 

\usepackage{textcomp} 
\usepackage[hidelinks]{hyperref} 
\usepackage{color}
\usepackage{printlen} 
\usepackage{float}
\usepackage{upgreek}
\usepackage{placeins}

\usepackage{xspace}

\author{Brendan P. Dyett}
\email{brendan.dyett@rmit.edu.au}                 
\affiliation{School of Science, RMIT University, Melbourne, VIC 3001, Australia}
\author{Xuehua Zhang}   
\email{xuehua.zhang@ualberta.ca}     
\affiliation{Department of Chemical $\&$ Materials Engineering, University of Alberta, Edmonton, T6G1H9, Alberta, Canada}       

\title{Accelerated Formation of H$_2$ Nanobubbles from a Surface Nanodroplet Reaction}


\begin{document}
	
	\begin{tocentry}
		\centering
		\includegraphics[trim={0 0 0 0}, clip,width=1\textwidth]{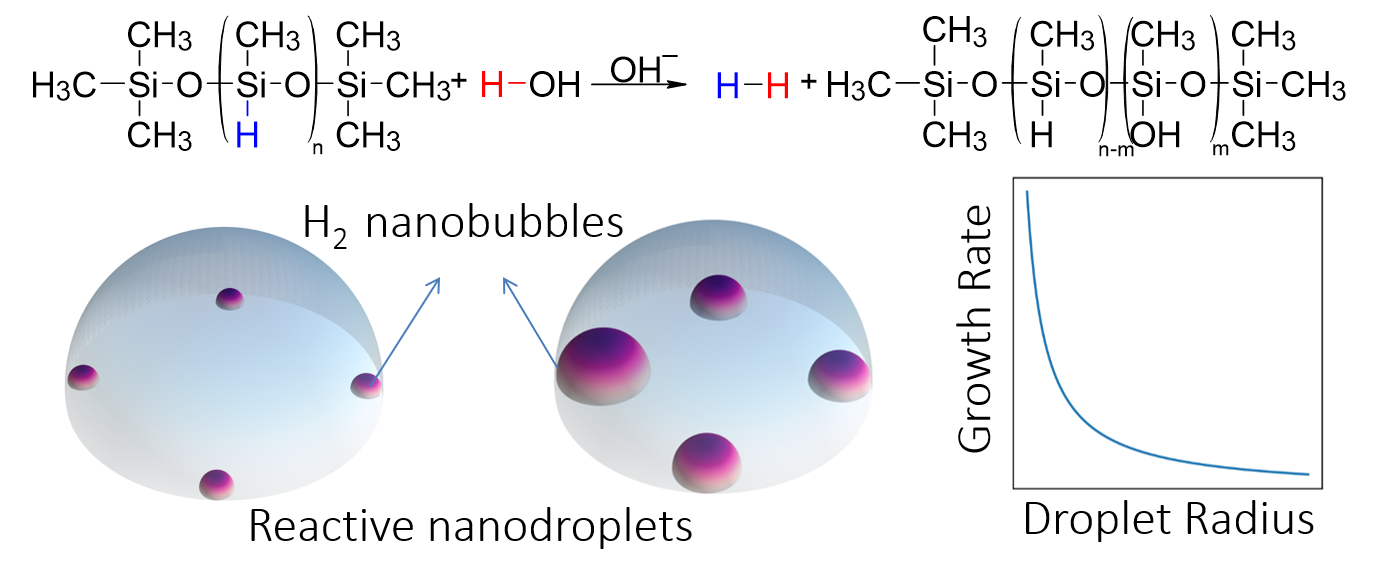}
		
	\end{tocentry}
	
	\begin{abstract} 

The compartmentalization of chemical reactions within droplets has advantages in low costs, reduced consumption of reagents and increased throughput. \textcolor{black}{Reactions} in small droplets have also been shown to greatly accelerate the rate of many chemical reactions. The accelerated growth rate of nanobubbles from nanodroplet reactions is demonstrated in this work. The gaseous products from the reaction at the nanodroplet surface promoted nucleation of hydrogen nanobubbles within multiple organic liquid nanodroplets. The nanobubbles were confined within the droplets and selectively grew and collapsed at the droplet perimeter, as visualized by microscopy with high spatial and temporal resolutions.  The growth rate of the bubbles was significantly accelerated within small droplets and scaled inversely with droplet radius. The acceleration was attributed to confinement from the droplet volume and effect from the surface area on the interfacial chemical reaction for gas production. The results of this study provide further understanding for applications in droplet enhanced production of nanobubbles and the on-demand liberation of hydrogen.

	\end{abstract}
	

\section{}
Droplets readily facilitate the miniaturization and compartmentalization of reactions as `mini reactors'. Their discrete nature lends itself to cost-effective, high throughput screening and sensing applications.\cite{kelly2007miniaturizing,wang2009efficient,zhang2017ionic,guardingo2016reactions,zhu2008lab, feng2018droplet,dittrich2006lab,ueda2012dropletmicroarray}
Many chemical reactions confined within droplets may also occur much faster than in bulk, in some cases, by a factor of a million.\cite{bain2016accelerated,yan2016organic,li2016role,banerjee2017can,yan2017two,kuksenok2014chemical,girod2011accelerated} Similar acceleration has also been demonstrated for reactions taking place within a thin liquid film.\cite{wei2017reaction} 
The approach draws parallels to `on-water' chemistry, where the rate of reaction and product collection are enhanced at the water interface, despite of poor reagent solubility.\cite{narayan2005water} Now, so called `on-droplet' chemistry is attracting increasing research attention to accelerate reactions even further.\cite{bain2017droplet}

The mechanisms of acceleration are still under investigation, however, are generally attributed to the following factors. First, the concentration of reagents within droplets may be progressively increased as the solvent component evaporates; as is the case in sprays during mass-spectrometry. Second,  the diffusion of reagents is enhanced within small droplets. Moreover, the collision and mixing of droplets may enhance the rate by further reducing the effective diffusion length scale.\cite{carroll2013experimental,lee2015microdroplet,davis2017colliding} Both these influences are intuitively rationalized through concepts such as collision theory. 

Arguably the most intriguingly and complex factor is influence from the droplet interface. Several studies have demonstrated that the interface itself may alter the reaction.  Nakatani and coworkers utilized optical trapping to demonstrate that dye formation and electron transfer were accelerated at singular droplet interfaces. The enhancement was attributed to the increasing surface area to volume ratio as the droplet size was reduced.\cite{nakatani1995droplet,nakatani1995direct,nakatani1996electrochemical}
Meanwhile, Fallah-Araghi \textcolor{black}{\textit{et al.}} demonstrated that reaction rate was inversely proportional to the droplet radius owing to the favourable  adsorption/desorption of products at the droplet interface, shifting equilibrium forward.\cite{fallah2014enhanced}
Li \textcolor{black}{\textit{et al.}} further demonstrated that the transport of precursors to the interface was proportional to the reaction rate and could be varied according the the reagent substituents.\cite{li2016role} 
Besides altering the reaction rate, microdroplet reactions sometimes yield products different from their bulk counterparts or occur \textit{via} distinct intermediates.\cite{baffou2014super,nam2017abiotic}
This scale influence on the reaction is so pronounced that droplet mediated reactions are a burgeoning area of interest within prebiotic chemistry as a component to the origins of life.\cite{vaida2017prebiotic,urban2014compartmentalised} 
Nam \textcolor{black}{\textit{et al.}} showed that the phosphorylation of sugars occurred spontaneously in aqueous microdroplets. The entropic barrier observed in bulk solution was found to be negligible within droplets, suggesting that reduced entropy change may be attributed to the alignment of reactants at the interface. \cite{nam2017abiotic} At low concentrations, various biomolecules such as pyruvate and cysteine have been shown to spontaneously reduce at the droplet-air interface. \cite{lee2018spontaneous_bio} Moreover, it was recently demonstrated that gold nanowires could be prepared within microdroplets despite the absence of reducing agents.\cite{lee2018spontaneous} Banerjee and Zare have also highlighted that surface charge, namely, surface protons of the droplet, could replace the need for acid catalysts.\cite{banerjee2015syntheses} 
These discoveries demonstrate the replacement of a pivotal reagents by the physical properties of the interface, alluding to an alternative approach for enhanced chemical reactions. \cite{nam2017abiotic,lee2018spontaneous_bio,banerjee2015syntheses,lee2015acceleration} 

In studying reactions confined by liquid-air interfaces, microdroplets are most often produced as atomized sprays, such as in electrospray ionisation (ESI). The sprays yield a range of droplet radii typically between 1 - 50 $\upmu$m. The various adaptations of this method have recently been reviewed.\cite{ingram2016going} During flight, reactions take place as the droplets evaporate, before the product(s) are collected, typically by a mass spectrometer inlet. Multiple droplet sources may also be combined to promote collisions.\cite{lee2015microdroplet} The distance, velocity of the spray can be experimentally controlled or subjected to other experimental conditions such as light.\cite{chen2016picomole} In an effort to accurately control droplet size, piezoelectric dispensers were recently utilized to study reactions within singular droplets.\cite{jacobs2017exploring} 
\textcolor{black}{At the same time, recent reports indicate challenges in disentangling the influence of compartmentalization from non-droplet related processes in ESI experiments. Gallo \textit{et al.}\cite{gallo2019chemical} suggested the results observed in electrosprays were not necessarily reflective of pristine droplet-air interfaces. 
Meanwhile Jacobs \textit{et al.}\cite{jacobs2018studying} demonstrated gas-phase reactions could also account for the products attributed to in-droplet processes. These findings suggest the development of alternative experiment designs are necessary in advancing the field. 
Alternative methods including} larger sized Leidenfrost droplets from simple pipette deposition has also been studied.\cite{bain2016accelerated}
Reactions confined within liquid-liquid interfaces have also been investigated by multiple droplet sprays.\cite{yan2017two} In addition, several studies have utilized emulsions (radius $\sim$ 1 - 30 $\upmu$m) to study reactions internal to the droplet\cite{fallah2014enhanced}  and between the droplet and bulk phase.\cite{nakatani1995droplet,nakatani1995direct,nakatani1996electrochemical,zhang2017ionic}

In terms of further elucidating the mechanisms of droplet accelerated reactions, surface \textcolor{black}{nanodroplets} may provide many advantages. 
Surface nanodroplets are droplets with lateral diameters between 0.1 – 10 $\upmu$m, heights ranging 10 – 1000 nm, and typical volume of the order femto- or attoliters; (10$^{-15}$ L, 10$^{-18}$ L, respectively.) Solvent exchange demonstrates\textcolor{black}{} flexibility in terms of controlling droplet size, morphology and liquid properties.\cite{Zhang2015,dyett2017formation} Solvent exchange describes the heterogeneous nucleation and growth from mixing induced oversaturation. The droplet size can be tailored by solution and flow conditions while the droplet morphology can be tailored by the relevant interfacial tensions.\cite{Zhang2015,Bao2015,Lu2016} Different compositions in the droplets can be formed by co-precipitation from the mixture of the solutions. \cite{li2018formation}   Surface nanodroplets are stabilized by pinning effect on the boundary. Their long-term stability allows for following the reaction dynamics with high temporal and spatial resolutions.

Despite of significant influences on a wide range of chemical reactions, formation and growth of nanobubbles from droplet reactions has been rarely explored.  The formation of bubbles confined within fluids is in and of itself relevant to a number of processes,\cite{hain2019multimodal} such as food and material manufacturing, and local heating in thermal therapy by plasmonic effect of nanoparticle.\cite{vincent2017statics,dyett2019} In particular, the formation of bubbles within confined \textit{nano/micro}-fluids is a growing field which has largely only been discussed theoretically.\cite{favelukis2004dynamics,xu2010numerical,arefmanesh1991diffusion,doinikov2018natural,doinikov2018cavitation,doinikov2018model}  Experimentally, the cavitation of bubbles has been explored in droplets of the order $\sim$ 20-150 $\upmu$m.\cite{vincent2012birth,vincent2014fast,shang2016osmotic} It is unclear how the droplet size influences the evolution of gas nanobubbles produced from a droplet reaction. 

Here, we exploit the unique properties of surface nanodroplets to expand the portfolio of droplet mediated reactions. While existing studies have been isolated to liquid phase reactions, we will demonstrate the liberation of gaseous products as surface nanobubbles confined by surface nanodroplets. 
 Gas-evolution reaction with polymeric and composite surface nanodroplets liberates $H_2$ gas and subsequently drives the formation of nanobubbles ($<$ 200 nm). By following the size of nanobubbles, the effect of the droplet size on the growth rate of nanobubbles is unravelled.  The findings from this study highlight the opportunities for enhanced nanobubble formation and the on-demand liberation of hydrogen as clean energy source by control and design of the size of reactive droplets.

\begin{figure*}[htp]
	\centering

	\includegraphics[trim={0 6.8cm 0 0}, clip,width=0.95\textwidth]{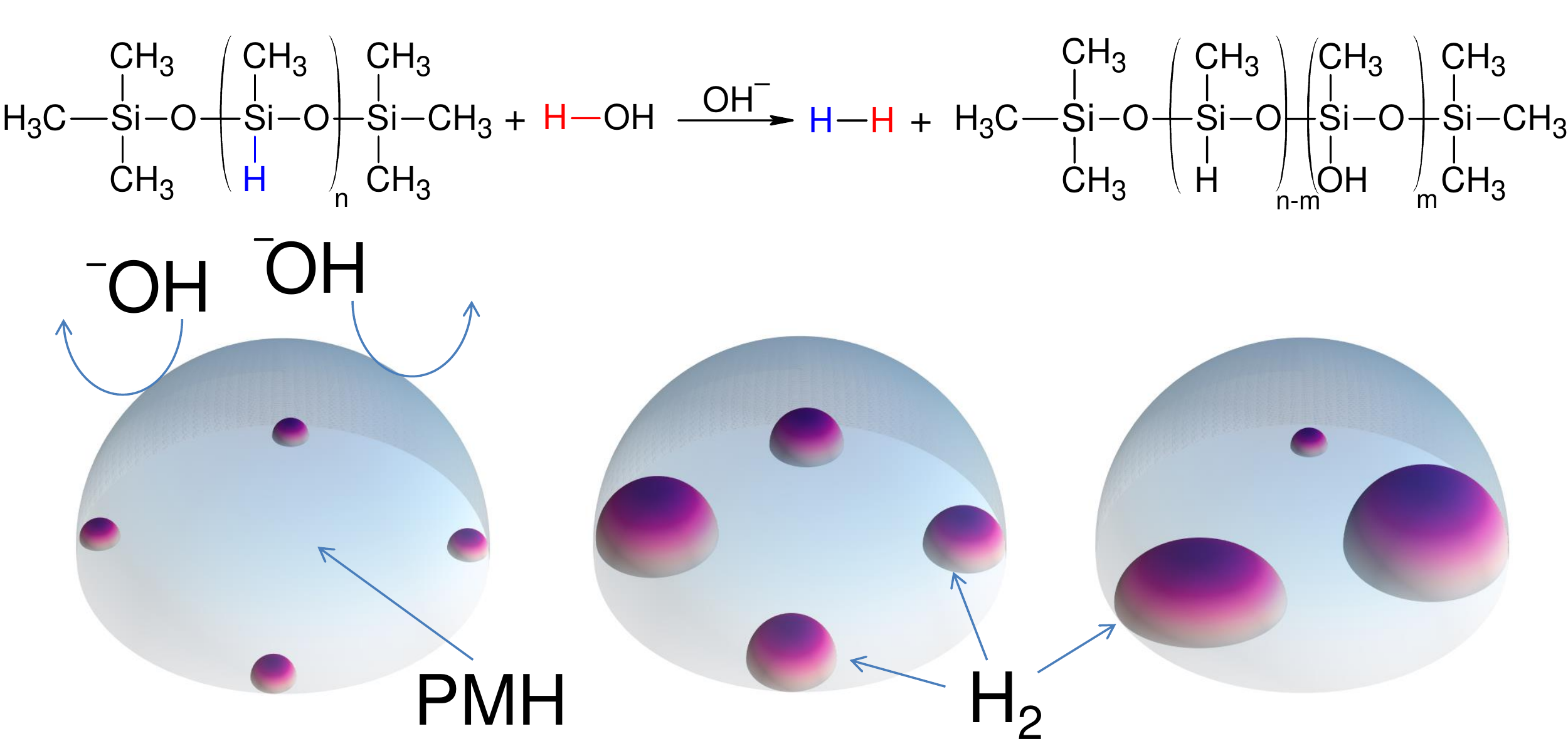}
	\caption{NaOH catalyzed dehydrogenation of PMH. The two-phase reaction between PMH microdroplets and surrounding aqueous media is depicted schematically. The droplets are shown as the larger blue spherical caps, with the smaller hydrogen bubbles shown as the purple spherical caps.}
	\label{reaction_scheme}
\end{figure*}

\section{Results and Discussion} 

\textbf{Nanobubble formation from droplet reaction }

Hydrogen nanobubbles were produced \textcolor{black}{from} the biphasic reaction between PMH nanodroplets and water, catalyzed by NaOH in the solution as shown in Figure \ref{reaction_scheme}. PMH is attacked by hydroxide ions, driving the liberation of H$_2$, the reformation of hydroxide and the progressive formation of a copolymer, poly(methylhydro-co-methylhydroxy)siloxane. \cite{toutov2016sodium}  The reaction takes place at the surface of PMH droplets as water is insoluble in PMH droplets.

\begin{figure*}[htp]
	\centering
	\includegraphics[trim={0 11cm 8cm 0}, clip,width=1\textwidth]{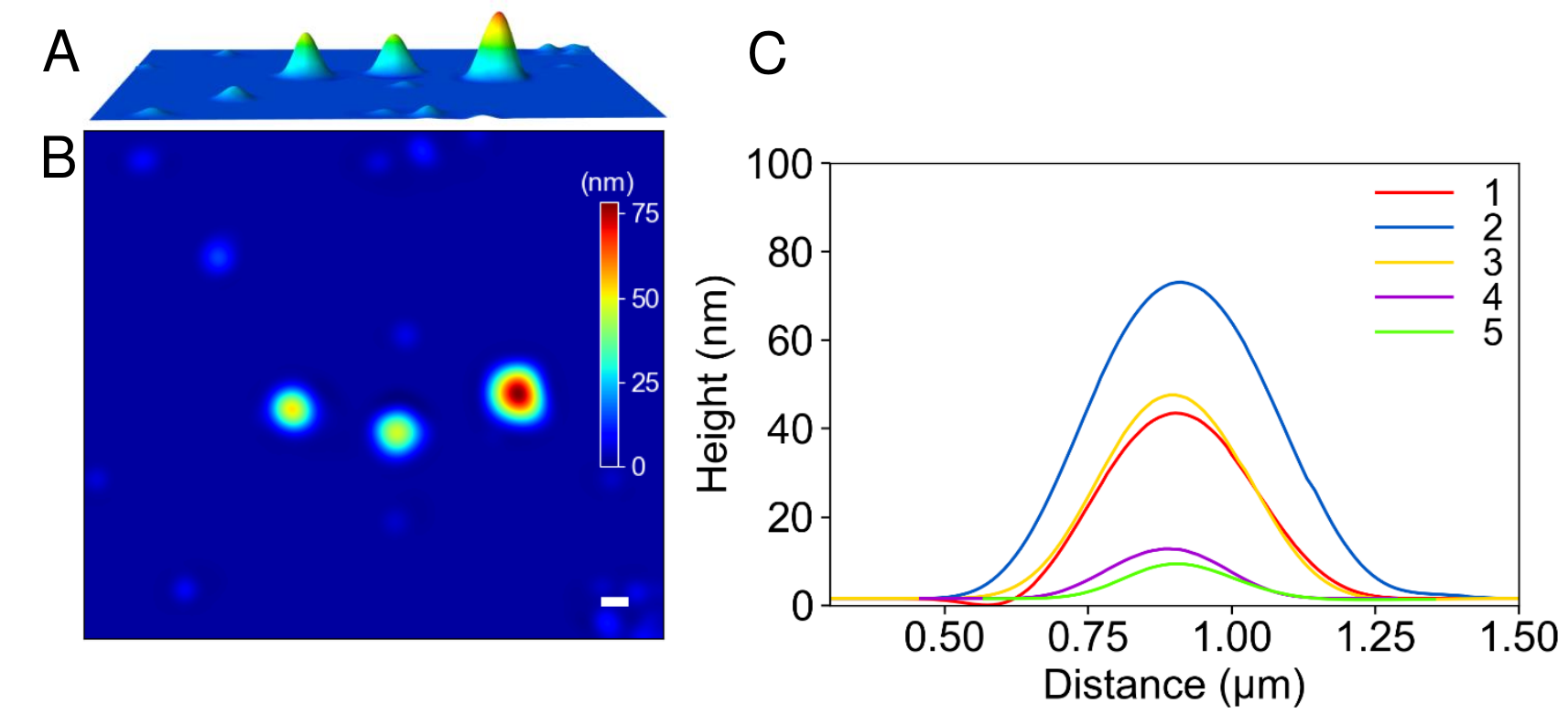}
	\caption{(A) 3D Height topology of immersed PMH surface droplets.  (B) Corresponding 2D height profile of immersed PMH droplets. Scale bar = 200 nm. (C) Extracted height profile for 5 representative droplets. } 
	\label{AFM}
\end{figure*}

 The size of the droplets produced by solvent exchange was heterogeneous, enabling comparison of the effect of droplet size on bubble production rate in the same experiment. The morphology of PMH droplets was revealed by liquid AFM, as shown in Figure \ref{AFM}. These droplets exhibit and maintain spherical cap geometry from a range of droplet sizes with lateral diameters $\sim$ 250 $nm$ - 1 $\upmu$m.
 For droplets with  0.7 $\upmu$m $<$ $D$ $<$ 1 $\upmu$m the contact angle was measured by height imaging in water to be $\sim$ 14-18$^\circ$, possibly influenced by slight deformation from AFM imaging.\cite{zhang2008interfacial} 
 Utilizing the PeakForce imaging at an extremely low set-point force gave the contact angle of the droplets on the order $\sim$ 20 $\pm$ 2 $^\circ$ in water, still smaller than  the macroscopic contact angles in Table \ref{Table1}. This result is consistent with the previous report that the contact angle of nanodroplets is generally smaller than the macroscopic counterparts. \cite{Zhang2015}

\begin{figure*}[htp]
	\centering
	\includegraphics[trim={0 6.8cm 0 0}, clip,width=0.95\textwidth]{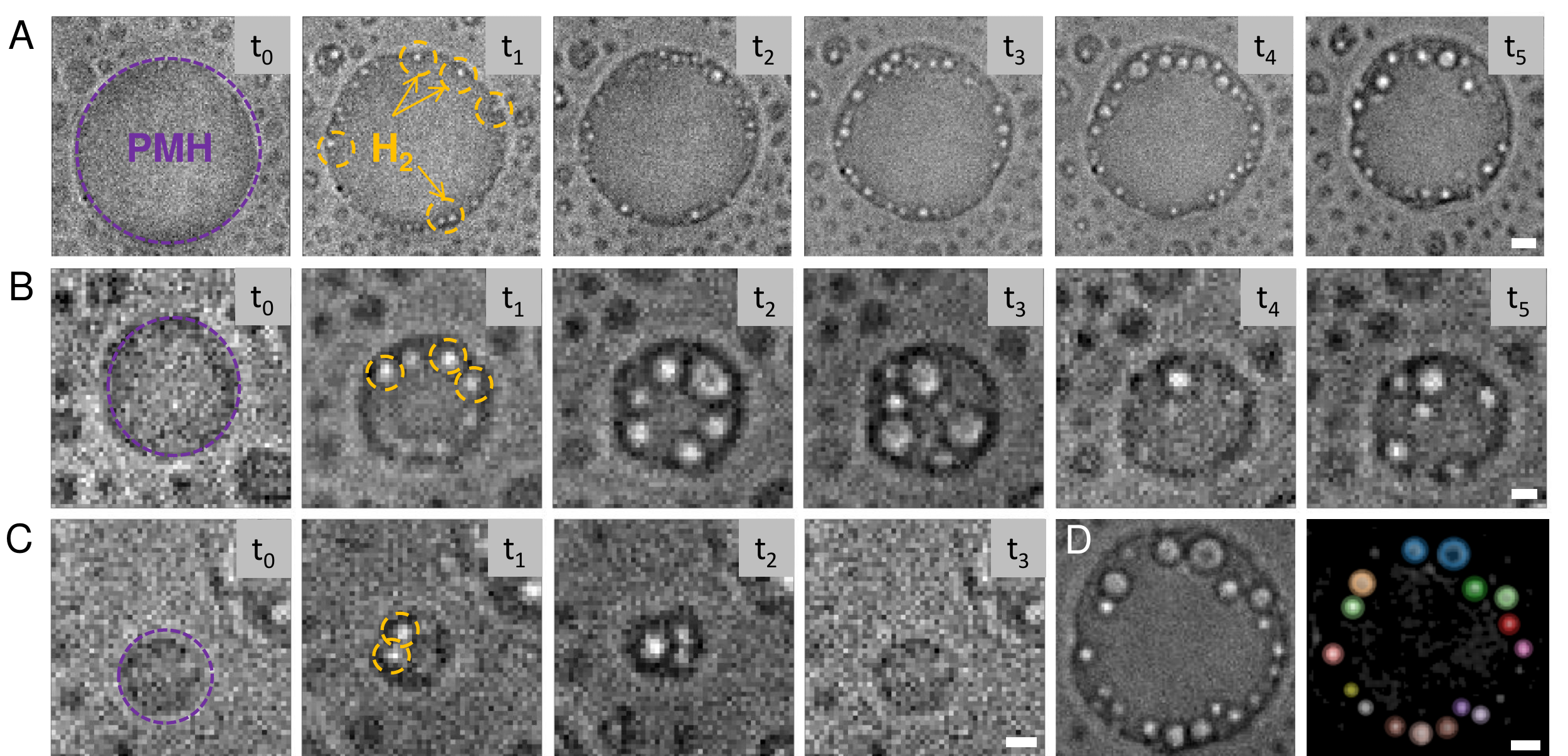}
	\caption{(A) Series of optical images of $H_2$ nanobubbles produced from a PMH droplet (D $\sim$ 15 $\upmu$m). The large initial droplet is highlighted in purple while the formed $H_2$ bubbles are highlighted in orange. Over time numerous bubbles were observed to form over the perimeter of the droplet. The relative times for $t_{0-5}$ are 0 s, 320 s, 640 s, 960 s, 1280 s and 2560 s. Scale bar = 2 $\upmu$m.  (B) Series of optical images of $H_2$ nanobubbles produced from a PMH droplet (D $\sim$ 6 $\upmu$m). The relative times for $t_{0-5}$ are 0 s, 320 s, 480 s, 640 s, 960 s and 1120 s. (C) Series of optical images of $H_2$ nanobubbles from a PMH droplet (D $\sim$ 2.5 $\upmu$m). The relative times for $t_{0-3}$ are 0 s, 160 s, 256 s and 365 s. Scale bar = 1 $\upmu$m. The concentration of NaOH is 0.048 M, same for (A)-(C).  (D) Representative optical image prior to and following image processing. Individual bubbles are highlighted in assorted colours. Scale bar = 2 $\upmu$m.}
	\label{bubble_images}
\end{figure*}

The bubble formation was induced by addition of sodium hydroxide solution. Certain incubation time was required before onset of nanobubble formation. 
Three representative examples of the visualized reactions in varying droplet size are shown in Figure \ref{bubble_images}. 
A largest PMH droplet with a diameter $D$ of $\sim$ 15 $\upmu$m is shown in Figure \ref{bubble_images}A. At $t_1$ $\sim$ 320 s, a number of small bubbles can be seen as bright circular regions at the perimeter of the droplet. The bubbles were approximately $\sim$ 500 $nm$ in diameter.  Over time, an increased number of bubbles formed and grew. 
Next series in Figure \ref{bubble_images}B shows the case of a smaller droplet with $\sim$ 6 $\upmu$m in diameter. Here it is clear that the bubbles could grow ($t_1$ – $t_2$), coalesce ($t_2$ – $t_3$) and collapse ($t_3$ – $t_4$). The final frame demonstrates that following bubble collapse, new bubbles nucleated around the vacated areas of the rim.  Figure \ref{bubble_images}C shows the series of the smallest droplet with a diameter $D$ of only $\sim$ 2.5 $\upmu$m. \textcolor{black}{For} this series, multiple bubbles are already observed at $t_1$ $\sim$ 160 s. A full cycle of bubble growth and collapse across the droplet was completed within 365 s.  

From the collective series, the bubbles showed \textcolor{black}{a} tendency to nucleate and grow at the droplet rim. This is in stark contrast to cavitation experiments which show highly localized formation.\cite{vincent2014fast,shang2016osmotic}
The bubbles did not appear to occupy the inner portions of the droplet until a significantly later stage of the reaction. Given the droplets are small,  the product diffusion would be rapid throughout the droplet. This result may reflect a reduced nucleation barrier at the droplet rim, analogous to capillary condensation. The other possibility is that the reaction rate is much faster around the droplet rim, in analogy to coffee stain effect in droplet evaporation. \cite{deeganPRE} \textcolor{black}{The} bubble size remained consistent throughout a range of droplet sizes. 
It is observed and rationalized through geometric confinement that fewer bubbles fit along the droplet perimeter as the droplet size reduces. In addition, the bubbles were forced to occupy the increasing proportion of the droplet.

\begin{figure}[htp]
	\centering
	\includegraphics[trim={0 0 0 0}, clip,width=0.475\textwidth]{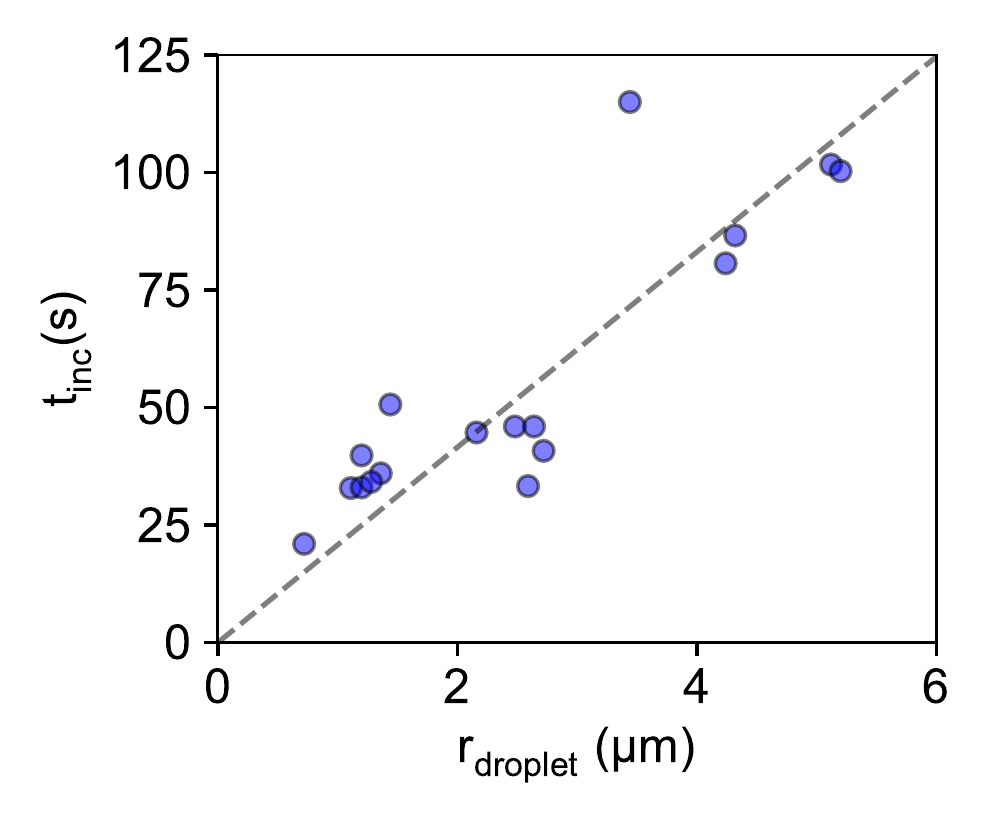}
	\caption{Plot of the observed incubation time $t_{inc}$ \textcolor{black}{\textit{vs}} the radius $r_{droplet}$ of the reacting droplet, derived from the first bubble detected in the images. \textcolor{black}{The dashed black line indicates a linear fit as a guide.}}
	\label{pmh_inc}
\end{figure}

\textcolor{black}{The} bubbles appeared earlier in smaller droplets, as shown in Figure \ref{pmh_inc}. \textcolor{black}{Classical nucleation theory typically dictates a non-linear relationship for the probability of bubble nucleation, which in this case, is complicated by both droplet geometry and chemical reaction. Here, a linear slope is shown to indicate the general trend over this window of droplet size.} 
Given that H$_2$ is a product of the reaction, with time the reaction should drive oversaturation within the droplet and eventually induce nucleation and growth of surface bubbles. Moreover, the rate of bubble production should act as a proxy to track the reaction rate. 
Although the abundance of literature reports acceleration reaction rates in droplets\cite{fallah2014enhanced,nakatani1995droplet,banerjee2015syntheses,nam2017abiotic}, \textcolor{black}{our} results show that there is enhancement in the nanobubble production rate due to the reduction in the droplet size.   
\textcolor{black}{The}  bubble formation within such confined liquid domains demonstrates several intriguing features in this complex colloidal system. The subtleties in bubble dynamics will be discussed below. Then we will focus on the growth of singular bubbles, free of coalescence will be utilized to determine the dependence of the nanobubble growth rates on droplet size more quantitatively.

\textbf{Nanobubble dynamics}

\begin{figure*}[htp]
	\centering

		\includegraphics[clip,width=0.95\textwidth]{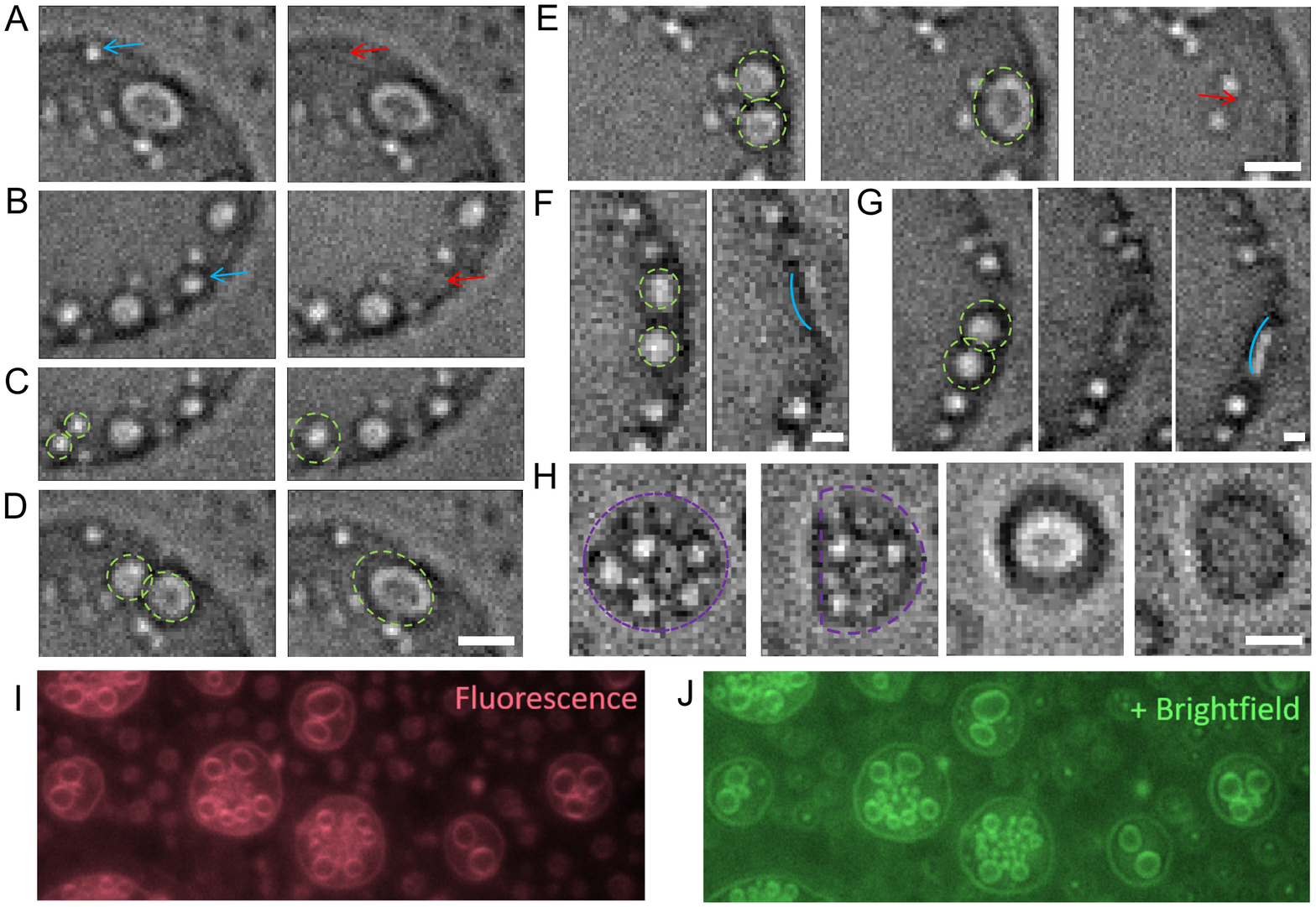}
	\caption{ Series of optical images highlighting bubble dynamics during growth. The presence and collapse of bubbles is highlighted by the blue and red arrows respectively.  Coalescence between bubbles is highlighted by dashed green circles. The change in morphology is guided by the coloured shapes. (A) Collapse of isolated bubble at droplet rim, bubble radius of 0.4 $\upmu$m. (B) Collapse of isolated bubble at droplet rim, bubble radius of $\sim$ 0.6 $\upmu$m. (C) Coalescence of two bubbles at the droplet rim, $\sim$ 0.3 $\upmu$m and 0.4  $\upmu$m, respectively. (D) Coalescence of larger bubbles,  bubble radius $\sim$ 0.7 $\upmu$m and 0.9 $\upmu$m, respectively. The final morphology becomes elliptical.  (E) Coalescence of larger bubbles,  radius of $\sim$ 0.9 $\upmu$m and 1 $\upmu$m, respectively. The final morphology becomes elliptical. After coalescence the bubble collapses. (F) Coalescence induced collapse. After the bubble collapse the morphology of the droplet altered.  (G) Coalescence induced collapse. After the bubble collapse the morphology of the droplet altered. (H) A number of bubbles form and coalesce triggering damage to the droplet morphology, highlighted in blue. Later, new bubbles form, coalescence and engulf the entirety of the droplet. Scale bar = 1 $\upmu$m. Fluorescence (I) and bright field (J) images of nanobubbles inside the droplets dyed by Rhodamine 6G. Image area: 50 $\mu$m by 20 $\mu$m. }
	\label{bubble_images_2}
\end{figure*}

The dynamical behaviours of nanobubbles around the droplet rim are captured with increased temporal resolution.  Figure \ref{bubble_images_2} highlights the appearance and disappearance of individual bubbles. Figure \ref{bubble_images_2}A-B show that the bubbles often collapsed with radii less than 0.5 $\upmu$m, before interacting with other bubbles. Conversely, in Figure \ref{bubble_images_2}C-E, some larger bubbles shown formed by coalescence. The coalescence of bubbles is indicated by dashed green circles. Following coalescence, bubble morphology may become elliptical in nature, indicating the initial bubbles were strongly pinned.  Figure \ref{bubble_images_2}E captures the process that following coalescence, the bubbles may then collapse. It is expected that the bubble collapse was induced by rupture of the thin polymer film between the gaseous and aqueous phase. Upon this rupture, a significant increase in Laplace pressure would be expected. Here interfacial tension $\gamma_{pmh-air}$ is $\sim$ 20 mN/m and $\gamma_{water-air}$ is $\sim$ 72 mN/m. The collapse of nanobubbles may induce significant damage to underlying materials.\cite{zhang2014controlled} The droplet deformation is presented in \textcolor{black}{Figure} \ref{bubble_images_2}F-G. Following coalescence induced collapse, the droplet perimeter is seen to be locally concave.  A similar series is shown in \textcolor{black}{Figure} \ref{bubble_images_2}H, where the droplet perimeter is highlighted in purple. Following collapse, the droplet edge receded inwards, suggesting the bubble was occupying the space in the droplet. The later frames of this series demonstrate that bubbles may coalescence into a bubble that becomes large enough to engulf and detach the droplet. The bright and fluorescence images in \textcolor{black}{Figure} \ref{bubble_images_2}I,J clearly show that inside a reacting droplet, larger nanobubbles are near the rim while small nanobubbles are forming from the inner area. 

How fast nanobubbles evolve with time is closely related to the reaction conditions. At a high concentration of NaOH, dramatic differences in nanobubble dynamics were observed. After addition of NaOH, the droplets appeared to rapidly reduce in size and detach from the substrate. It is suspected that the rapid production of gas, as well as the bubble collapse, provides sufficient force to detach the droplets from the substrate. Representative images of this process are shown in Figure S1. The result at higher concentration of NaOH  is indicative of further increases to the reaction rate, although the exact rate was very difficult to measure in our current experimental setup. These results indicate that the bubble formation at the rim only occurs within a range of reaction conditions with a mild level of hydrogen oversaturation.

\textbf{Dependence of nanobubble growth rate on droplet size}

\begin{figure}[htp]
	\centering
	\includegraphics[trim={0 0 0 0}, clip,width=0.95\textwidth]{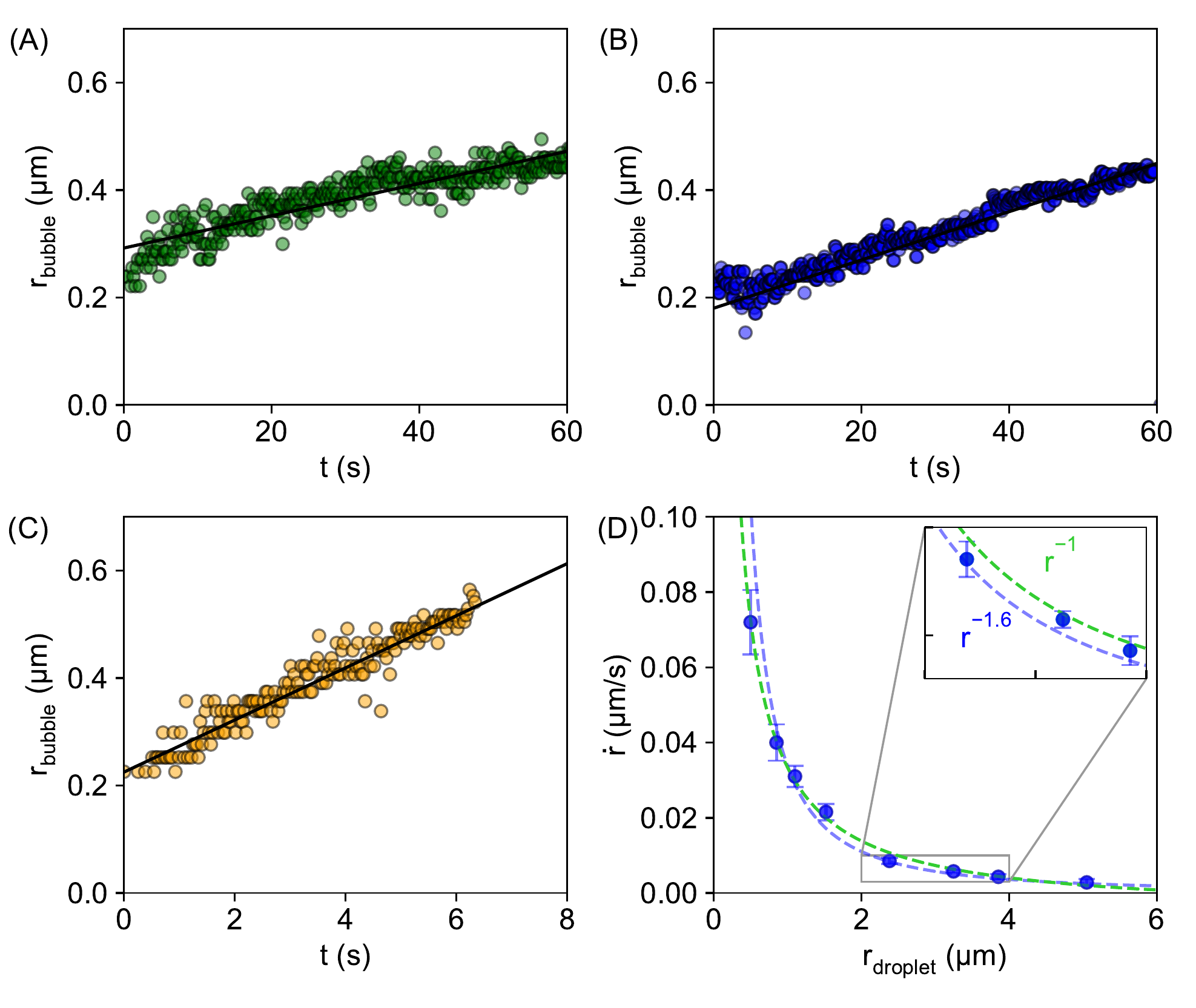}
	\caption{Plot of the radius of a representative individual bubble \textcolor{black}{\textit{vs}} time, formed in a droplet with the radius $\sim$ 5 $\upmu$m in (A), 3 $\upmu$m in (B) and 1 $\upmu$m in (C), respectively.  The corresponding linear best fits yield growth rates of $\sim$ 0.0031 $\upmu$ms$^{-1}$, 0.0045 $\upmu$ms$^{-1}$ and 0.049 $\upmu$ms$^{-1}$. (D) Plot of the average bubble growth rate \textcolor{black}{\textit{vs}} droplet size at [NaOH] = 0.048 M with the best exponent fit of -1.6 in blue and enforced exponent of -1 in green.  }
	\label{pmh_size}
\end{figure}

To further investigate the correlations between reaction conditions and production of nanobubbles, the growth rate of bubbles in various droplet sizes were determined.  \textcolor{black}{The droplet size is defined as the initial droplet radius before reaction and any of the observed changes shown in Figure \ref{bubble_images} \& \ref{bubble_images_2}. The change in droplet size due to the reaction is presented in Figure S2 and shows a negligible change within the growth time of each bubble.} The nanobubble growth was determined by particle tracking as shown in processed image in Figure \ref{bubble_images}D, following the same procedure in our recent work \cite{dyett2019}. Due to the resolution and noise of the measurements, linear fits were utilised to quantify the growth rate. \textcolor{black}{Further, we find that due to the collapse of bubbles, the window of their growth is well described by linear slopes, as presented in Figure S3.} Figure \ref{pmh_size}A-C, shows a representative bubble growth as radius \textit{vs} time for bubbles formed in a droplet with the radius $\sim$ 5, 3 and 1 $\upmu$m, respectively. From the plots, the radius of nanobubbles steadily increased until a final radius of $\sim$ 500 $nm$. Notably, the time taken to reach the maximal radius progressively decreases with reduction in the droplet size, yielding growth rates of $\sim$ 3.1 nms$^{-1}$, 4.5 nms$^{-1}$ and 49 nms$^{-1}$, respectively. 
The average growth rate as a function of the droplet size was determined and plotted in Figure \ref{pmh_size}D. \textcolor{black}{The} average growth rate revealed a strong dependency on droplet size, namely, that the bubble growth rate was accelerated in smaller droplets. The best exponential fit, demonstrates rate scales as rate $\sim$ $r_d ^{-1.6}$. That said, ${r_d}^{-1}$ can also fit the experimental closely within the accuracy of our experimental measurements.

\begin{figure}[htp]
	\centering
	\includegraphics[trim={0 0 0 0}, clip,width=0.95\textwidth]{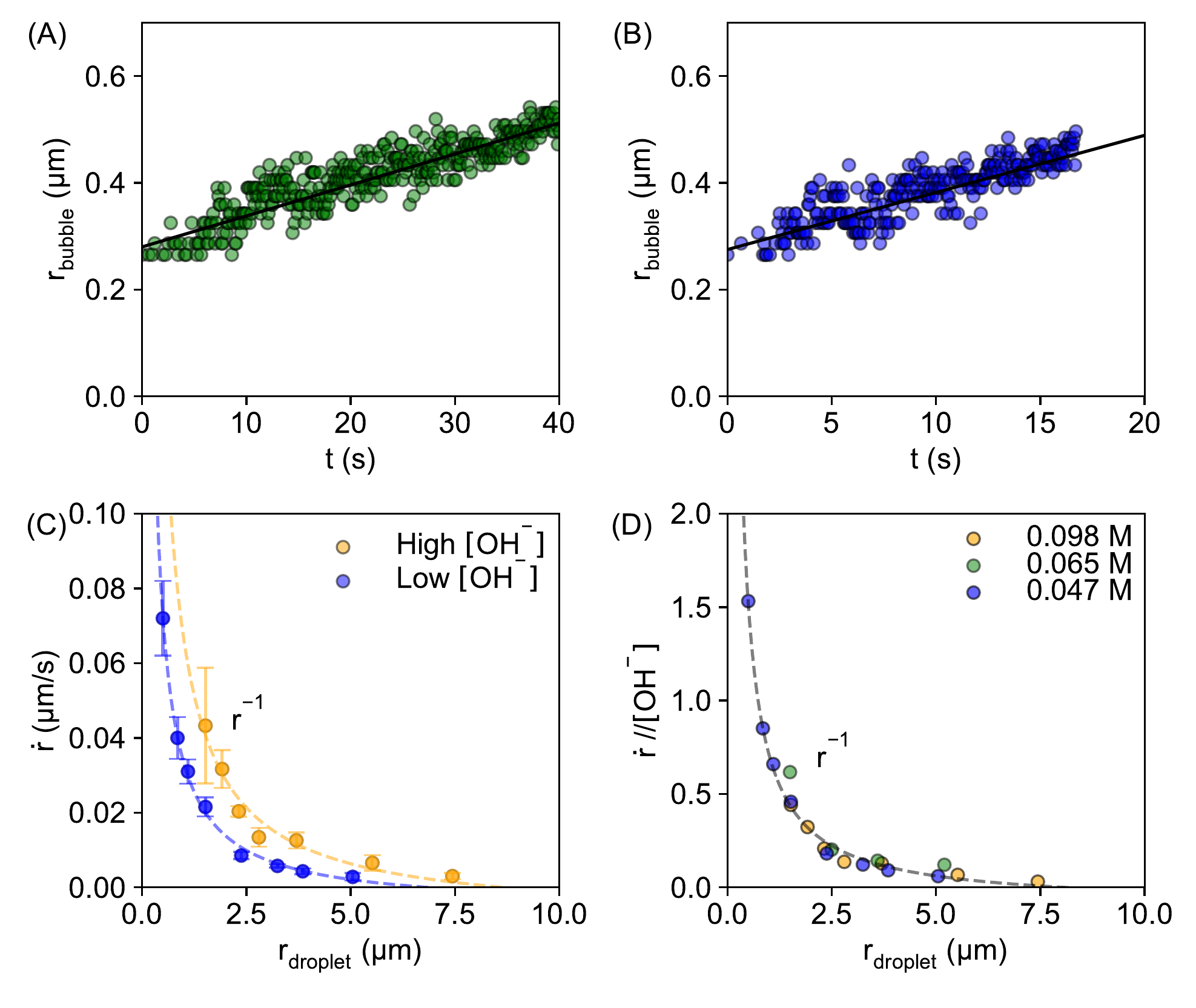}
	\caption{(A-B) Representative plot of an individual bubble growth as radius \textcolor{black}{\textit{vs}} time, formed in a droplet with the radius $\sim$ 6 $\upmu$m  and $\sim$ 4 $\upmu$m with [NaOH] of 0.098 M in the surrounding solution.  The corresponding linear best fits yield growth rates of $\sim$ 0.0059 $\upmu$ms$^{-1}$ and $\sim$ 0.011 $\upmu$ms$^{-1}$.  (C) Plot of the average bubble growth rate \textcolor{black}{\textit{vs}} droplet size at high and low [NaOH] of 0.098 M (orange) and 0.047 M (blue). The exponents for the orange and blue fittings of n = -1.   (D) Plot of bubble growth rate \textcolor{black}{\textit{vs}} droplet size normalized by sodium hydroxide concentration. 0.098 M, 0.065 M, and 0.048 M are plotted as orange, green and blue respectively. }
	\label{pmh_ph}
\end{figure}

To investigate the dependency of bubble growth rate on the concentration of reagents, the concentration of [NaOH] in the surrounding medium was increased two fold to 0.098 M.  Representative plots of bubble radius with time are shown in Figure \ref{pmh_ph}A-B. The growth rate was determined to $\sim$ 5.9 nms$^{-1}$ and $\sim$ 11 nms$^{-1}$ within a droplet of radius approximately 6 $\upmu$m and \textcolor{black}{4} $\upmu$m, respectively.  The reaction rate increased by more than two fold, compared to Figure \ref{pmh_size}A-C indicate, as reasonably expected. Again, the growth rate was observed to be faster within smaller droplets. This increase in rate is again demonstrated in the plot of the average bubble growth rate \textcolor{black}{\textit{vs}} droplet size shown in Figure \ref{pmh_ph}C. Once again the average bubble growth rate demonstrated a strong correlation to rate $\sim$ $r_d^ {-1}$ and best fit $r_d ^{-1.3}$. In both cases, the error bar is larger for smaller droplets, which is expected given the relative increase in spatial noise. 

Given that NaOH is acting as a homogeneous catalyst, the rate law may also be described as $Rate = k[PMH][OH^-]$. Then considering that aqueous volume is magnitudes larger than the volume of PMH, $[OH^-]$ will remain constant. It should then be expected that the reaction rate may be normalized by $[OH^-]$. The normalized rate is shown in Figure \ref{pmh_ph}D, showing an excellent collapse between multiple concentrations. Obviously, the same collapse is achieved treating hydroxide as a reagent to the first order. The master curve highlights the universal dependance of the bubble growth rate with the droplet size as $1/r$. 

The accelerated reactions within pure droplets suggest that the accessibility of OH$^-$ to the droplet interface plays a determinant role within this reaction. Considering a typical Langmuir isotherm,\cite{langmuir1916constitution} the rate of adsorption of hydroxide to the interface can be described as $k_{ads}$[OH]$f$, where $f$ is the area fraction available for adsorption and is defined as $f$ = (1 - $\theta$), where $\theta$ describes the droplet surface area already occupied by droplet-OH intermediates. The results shown in Figure \ref{pmh_ph} indicate the adsorption of hydroxide to the interface was not saturated. Upon saturation, it would be expected that the rate become zero order with respect to [OH]. 

\begin{figure}[htp]
	\centering
	\includegraphics[trim={0 0 0 0}, clip,width=0.475\textwidth]{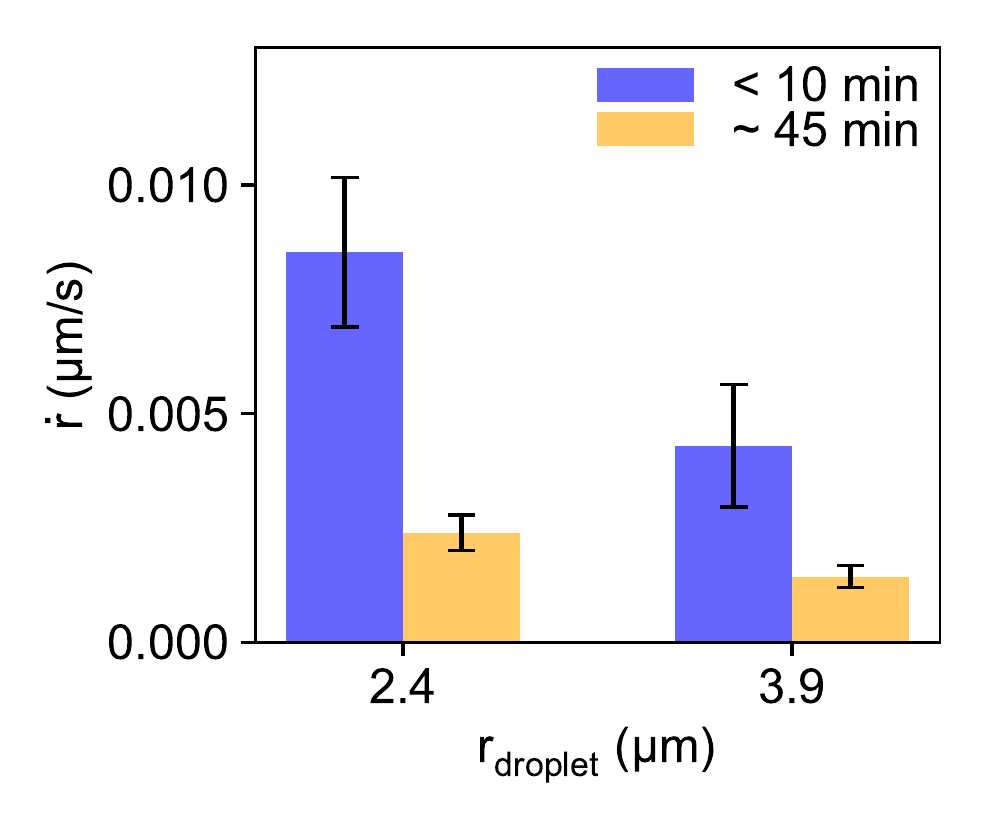}
	\caption{Plot of the average bubble growth rate for two droplet sizes at the early (blue) and late stage (orange). The concentration of NaOH in the surrounding solution is 0.048 M. }
	\label{pmh_dur}
\end{figure}

The experiments were typically recorded over a period of 60 – 90 min. From observations at later stages, it became evident that the growth rate of the bubbles dramatically slowed, reducing by a factor of $\sim$3 after $\sim$45 minutes due to a consumption of reagents and accumulation of the product as shown in Figure \ref{pmh_dur}.  After some time, it was also observed that the smallest droplets depleted themselves and no longer liberated bubbles. The reduction in reaction rate may be compounded by changes in the polymer within the droplet. During reaction the polymer chain is hydrolysed yielding reactive silanol groups. It is suspected that some crosslinking of the polymer may occur with sufficient time, impeding the reactivity of remaining functional groups.

\begin{figure*}[htp]
	\centering
	\includegraphics[trim={0 0 0 0}, clip,width=0.95\textwidth]{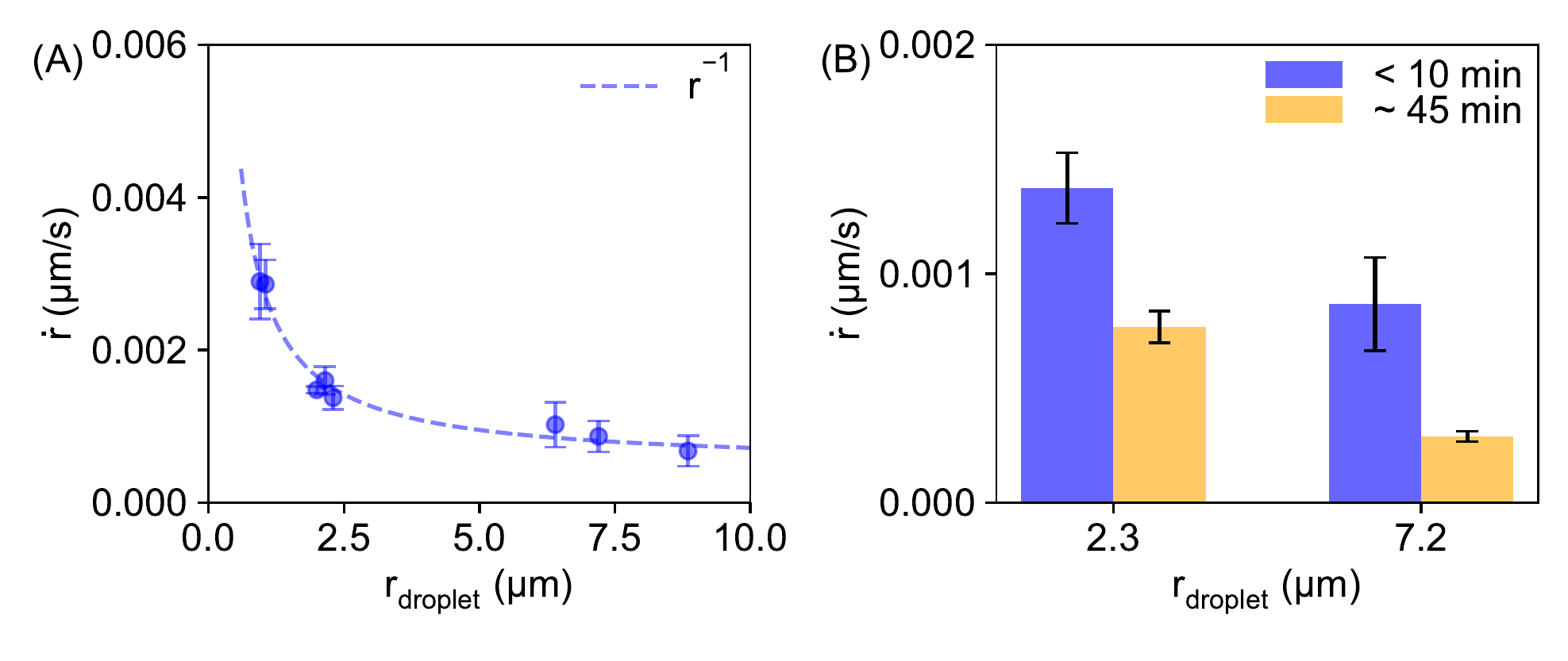}
	\caption{(A) Plot of the average bubble growth rate within composite decane-PMH droplets \textcolor{black}{\textit{vs}} droplet size at [NaOH] = 0.048 M. Fit to $r_d ^{-1}$ is highlighted in blue. Best fit was $r_d ^{-0.7}$ (B) Plot of the average bubble growth rate for two sizes of the composite decane-PMH droplet, at the early and late stage, shown in blue and orange respectively. }
	\label{dec_pmh}
\end{figure*}

To investigate the effect of PMH concentration on the reaction rate, solvent exchange was performed using a mixture of non-reactive oil decane and PMH. As shown in Figure \ref{dec_pmh}A, the reaction rate of the droplets consisting of decane and PMH again showed strong correlation to $r_d ^{-1}$. That said, the reaction of composite droplets revealed a reduced bubble growth rate, indicating the reaction was dependent to the concentration of PMH in the droplets. Furthermore, the coefficient $Q$ was obtained from fitting bubble growth rate by $\sim$ $Q r_d ^{-1}$, yielding the ratio $Q_{pmh}$/$Q_{composite}$ of $\sim$ 16. In other words, the reaction was $\sim$ 16 times slower for decane - PMH composite droplets. Based on our recent work\cite{li2018formation},  the composition of the surface droplets was determined to be $mass_{pmh}$/$mass_{dec}$  $\sim$ 0.073, corresponding to a $\sim$ 14 times dilution. Considering the reaction to be first order with respect to PMH, this dilution factor correlates well with reduced reaction rate.
From Figure \ref{dec_pmh}B, the reaction within the composite droplets also slowed with time, reducing by a factor of 2-3, after 40 minutes of reaction, \textcolor{black}{suggesting concentration of PMH within the droplet had also reduced by a factor 2-3.}.

\textbf{Proposed mechanism for enhanced nanobubble growth}

Our results under various conditions show that the bubble growth rate was all accelerated with reducing droplet radius within the range of $\sim$ 0.5 – 9 $\upmu$m.
A number of factors can contribute to the enhanced growth rate. Given that the volume of droplets was not reducing rapidly, the concentrating of reagents with decreasing droplet size as observed in electrosprays, can obviously be excluded. Our droplets can be regarded as a pure reagent, at least at the initial stage of hydrogen production. 

This dependence of the bubble growth rate on $r_d ^{-n}$ is in agreement with earlier reports where the increase in reaction rate increases with $r_d ^{-n}$, which was attributed to a thermodynamic advantage from the adsorption and desorption of species at the interface.\cite{fallah2014enhanced,li2016role} That said, there are important factors which differentiate the reaction here to those studies.  In those early reports, all reactants were internal to the droplet, whereas here the reaction is occurring between two phases, at the liquid-liquid interface. This biphasic reaction is also considered to be irreversible and never reach equilibrium. Considering stochastic nature of bubble nucleation, our experimental results show strong correlation between bubble growth rate and the droplet radius, expanding the scope of enhancement by reduced droplet size to nanobubble production.

The scaling of bubble growth rate with $r_d ^{-n}$ is rationalized as below. As the reaction is biphasic, taking place only at the surface of the droplet.
The rate can be expressed in terms of the interfacial area ($A_{droplet}$) as:

\begin{equation}
\frac{dm_{(g)}}{dt} = A_{droplet} * k * [PMH]_d [OH^-]_s
\end{equation}

Here 
the chemical reaction shown in Figure \ref{reaction_scheme} follows a second order rate law while $k$ is assumed to be a reaction constant.  None of [PMH], [$OH^-$]  or $k$ varies with the droplet size. The amount of hydrogen produced per unit time is proportional to the area of the droplet surface $A_{droplet}$ ($\sim r_{droplet}^2$). The produced gas from the reaction on the droplet surface diffuses out to increase the concentration of the dissolved hydrogen $C_g$ in the boundary layer adjacent to the droplet surface. During the period of incubation,  the concentration of hydrogen $C_g$ in the boundary layer increases with time. The liquid volume in the boundary layer is proportional to the droplet size, scaling with $r_{droplet}^3$. In per unit time the concentration of the dissolved hydrogen $C_g$ in the boundary layer increases with $A_{droplet}* k * [PMH]_d [OH^-]_s/V_{droplet}$, which leads to $C_g \sim r_{droplet}^{-1}$. 

Growth rate of a sphere bubble in an oversaturated environment was analytically calculated by Epstein and Plesset \cite{epstein1950stability}, and recently reviewed by Lohse and Zhang \cite{ZhangLohse2015}.  Assuming bubble growth is governed by the ideal gas, Young-Laplace and diffusion equations.\cite{ZhangLohse2015}, the radial growth rate

\begin{equation}
\frac{d r}{dt}= \alpha({\frac{1}{r} + \frac{1}{\sqrt{\pi Dt}}})
\label{diffusion}
\end{equation}

With

\begin{equation}
\alpha = \frac{D}{\rho_g} \frac{C_g - C_s (1+2\sigma/(rP_0)}{1+4\sigma/(3rP_0))}
\label{alpha}
\end{equation}

Here $D$ is gas diffusion coefficient, $\rho$ the density of the gas, $C_s$ the gas solubility, $\sigma$ the surface tension of the gas-liquid interface, $P_0$ the ambient pressure.  There are two limiting cases of big and small bubbles to simplify $\alpha$. \cite{ZhangLohse2015} For large bubbles, Laplace pressure inside the bubble is much smaller than the ambient pressure. For $r(t) \gg 2\sigma/P_0$, 
$\alpha \approx \frac{D (C_g - C_s) }{\rho_g}$.  Although the size of nanobubbles here is small, the condition may not necessarily satisfy $r(t) \ll 2\sigma/(P_0 C_g/C_s)$ because of $C_g/C_s \gg 1$. So the growth rate of these nanobubbles may be closer to the case of large bubbles.  Given the gas concentration $C_g \sim r_{droplet}^{-1}$, it is immediately clear that $\alpha$ and the growth rate of nanobubbles $\sim r_{droplet}^{-1}$.  The exact growth rate of hydrogen nanobubbles can be more complicated than a free bubble, due to geometrical derivation from a sphere, pinning effect from the substrate and the confinement from the droplet surface.

The above analysis would suggest that the bubble growth rate should decrease with time as the concentration of PMH decreases from depletion in the reaction at the droplet surface.  The initial growth rate of nanobubbles also decreases with the ratio of PMH to non-reactive liquid in the droplets, as observed experimentally. It is interesting to note that for surface droplets, as the contact angle and thus the surface-to-volume ratio may be tuned. This may prove desirable in future mechanistic or kinetic control studies for \textcolor{black}{nanobubble} growth rates.

\section{Conclusions}

Hydrogen nanobubbles produced from reactive polymer nanodroplets demonstrate an accelerated growth rate as the droplet radius is reduced. The bubbles demonstrated a preferential nucleation at the drop rim and revealed numerous growth dynamics including coalescence, collapse, and secondary nucleation. \textcolor{black}{It} was determined that the growth rate of the bubbles scaled inversely with the droplet radius. Furthermore, the reaction rate was proportional to both the concentration of reactants inside and outside the droplet. The experimental results simultaneously demonstrate the production of surface nanobubbles within pure reactive reagent and composite droplets. 
 The mechanism for the enhanced growth rate was the confinement of oversaturated gas in the droplet volume as the gas production from the interfacial reaction is proportional to the surface area of the droplet. The results invite future research toward droplet accelerated reaction, in particular, where the droplet may also contain a catalyst.  The findings from this study provide further understanding for applications in droplet enhanced chemical reactions and the on-demand liberation of hydrogen.

\section{Experimental method and simulations} 
\textbf{Chemicals and Materials} 

Polymethylhydrosiloxane (PMH) (Mn = 3400 g/mol, Sigma) and decane (99\%, Sigma) were utilized as to prepare reactive oils and were diluted in acetone (AR, Chem-supply) to prepare the first solution for solvent exchange. Sodium hydroxide (98\%, Sigma) and  Milli-Q water (18 M$\Omega$cm) were used to prepare basic solution to trigger the reaction.  8-well plates with bottom glass substrates (Lab-Tek, No. 1.5H) were used following chemical vapour deposition of 1H,1H,2H,2H-perfluorododecyltrichlorosilane (97\%, Sigma). The substrates were used immediately following sonication in Isopropanol (AR, Chem-supply) and drying by compressed nitrogen. All reagents were used as received without further purification.

\textbf{Droplet preparation and reaction conditions}

PMH droplets were prepared by solvent exchange within the well. The first solution was prepared as 1\% PMH in acetone (v/v), termed solution A. The second solution was Milli-Q water (18 M$\Omega$cm), termed solution B. Initially 0.2 ml of solution A was added to the well, followed by 0.6 ml of solution B. Upon addition of solution B, the solution was immediately cloudy, indicative of nanodroplet formation from ouzo effect. From the solution 0.2 ml was then removed and replaced by 0.2 ml of fresh solution B. This process was repeated 12 times to remove solution A thoroughly, then leaving a final well volume of 0.5 ml. PMH in solution A was removed in the experiments, reducing light scattering and allowed the reaction to be visualized within the surface droplets on the same plane with the glass substrate.

Composite nanodroplets consisting of PMH and decane were prepared by the solvent exchange in the well by following the same procedure for formation of PMH nanodroplets. Solution A was 0.91 vol. \% decane and 0.01 vol. \% PMH in acetone while solution B was water. Composite nanodroplets of PMH and 1-octanol were also prepared by the solvent exchange in the well by following the same procedure for formation of PMH nanodroplets. Solution A was 0.91 vol. \% 1-octanol and 0.01 \% PMH in acetone while solution B was 1-octanol saturated water. Both decane (99\%) and 1-octanol (99\%) were purchased from Sigma and used as received. 
The reaction for producing hydrogen nanobubbles was initiated by addition of the aqueous solution of NaOH into the well. The volume of the solution was 25 $\upmu$l to the concentration of the alkaline solution of $\sim$ 0.05 M in the well during the reaction. 

The physical parameters for the oil droplets are summarized in Table \ref{Table1}.  The viscosity $\mu$  of PMH is $\sim$ 45 cSt while the composite viscosity is estimated to be $\sim$ 4 cSt assuming an ideal mixture. The interfacial tensions with air and water demonstrated a downward trend with increasing PMH content. The macroscopic contact angle (CA) with air and water and the fluorinated substrate were $\sim$ 60$^\circ$ for both oil compositions.

\begin{table}[htp]
	
	\small
	\caption{Physical properties of PMH and the DEC-PMH mixtures at different volume ratios }
	\label{Table1}
	
	\begin{tabular*}{0.95\textwidth}{@{\extracolsep{\fill}}lllllll}
		\hline 
		Oil & $\gamma_{oil-water}$ & $\gamma_{oil-air}$  & CA in air & CA in H$_2$O& Density $\rho$ & Viscosity $\mu$ \\
		& $(mNm^{-1})$ & $(mNm^{-1})$ & $(^\circ)$ & $(^\circ)$ & $(kgm^{-3})$ & (mPa $\cdot$s)\\
		& & & & & &\\  
		\hline
		& & & & & &\\
		PMH & 0.4 & 19.9 & 60.7  & 60.5 & 1006 & 45\\
		DEC-PMH (91:9) & 30.8 & 24.6 & 59.2 & 58.6 & 760 & 4 \\ 
		DEC-PMH(50:50) & 25.6 & 22.3 & - & - & 868 & 23 \\ 
		\hline
	\end{tabular*}
	
\end{table}

\textbf{Characterization of hydrogen nanobubbles from droplet reaction}

The interfacial surface tensions and contact angles were determined by pendant and sessile droplet method using a goniometer (OCA20, DataPhysics Instruments GmbH, Germany). The interfacial tensions were determined using the `OpenDrop' package for python.\cite{OpenDrop2015}. Fluid AFM measurements were performed using an Asylum MFP-3D Bio AFM in AC mode with an iDrive cantilever holder and BL-TR400PB tip (nominal spring constant, $k$ = 0.02 N/m; freq. = 7-14 kHz). Additional Fluid AFM measurements were performed using a Dimension Icon utilizing ScanAsyst-Fluid+ tip (nominal spring constant, $k$ = 0.7 N/m; tip radius = 2 nm). The PeakForce setpoint was maintained during all images at 334.9 pN. 

The reaction experiments were recorded \textcolor{black}{\textit{in situ}} on a Nikon N-Storm super resolution confocal microscope (TIRF 100x, 1.49 NA objective lens).  A white-light LED was used to capture the surface of the substrate through bright-field imaging. The region of interest was collected by an Andor iXon DU-897 EMCCD camera, at 58 frames per second with a pixel calibration of 0.16 $\upmu$m per pixel. Image analysis was performed using a combination of NIS-Elements instrument software and homebuilt python codes, in part utilizing the open source PIMS, scikit\cite{van2014scikit} and TrackPy\cite{allan2016trackpy} packages for python. Each frame was enhanced by filtering and unsharp masking, before binarization. The bubbles were detected as bright, nominally circular features. In each frame, the size, shape and position of all detected bubbles were measured, as shown in Figure \ref{bubble_images}D. The linking of frames within allowed time and movement parameters allows for each bubble to be assigned a unique numerical identifier through time. \textcolor{black}{\textit{i.e.}} A bubble in the same location in subsequent frames is identified as the same bubble. Conversely, if the bubble collapses and is absent for a designated number of frames, any new bubbles in this location is given a new numerical identifier. The fluorescence images of reacting droplets were taken by using 488 nm laser as excitation. The droplets were binary mixture of PMH and decane, stained by Rhodamine 6G at \textcolor{black}{5} $\mu$M. The concentration of NaOH in the surrounding solution was 0.048 M.

\textcolor{black}{
\section{Supporting Information}
\textbf{Supporting Information Available:} (Figure S1) Time-lapse showing the PMH nanodroplet response to external stimuli of high concentration of NaOH. (Figure S2) Plot of droplet diameter \textit{vs} time during reaction. (Figure S3) Plot of the radius \textit{vs} time for representative bubbles with fiitting against equation 2.  This material is available free of charge \textit{via} the Internet at http://pubs.acs.org.
}
	
	\begin{acknowledgement}
	We thank the funding support from the Natural Sciences and Engineering Research Council of
Canada (NSERC) and Future Energy Systems (Canada First Research Excellence Fund). This research was undertaken, in part, thanks to funding from the Canada Research Chairs program.
 We also acknowledge the RMIT MicroNano Research Facility and the RMIT Microscopy and Microanalysis Facility for providing access to equipment and resources.
	\end{acknowledgement}
	
	
	\bibliography{paper}

\providecommand{\latin}[1]{#1}
\makeatletter
\providecommand{\doi}
  {\begingroup\let\do\@makeother\dospecials
  \catcode`\{=1 \catcode`\}=2 \doi@aux}
\providecommand{\doi@aux}[1]{\endgroup\texttt{#1}}
\makeatother
\providecommand*\mcitethebibliography{\thebibliography}
\csname @ifundefined\endcsname{endmcitethebibliography}
  {\let\endmcitethebibliography\endthebibliography}{}
\begin{mcitethebibliography}{66}
\providecommand*\natexlab[1]{#1}
\providecommand*\mciteSetBstSublistMode[1]{}
\providecommand*\mciteSetBstMaxWidthForm[2]{}
\providecommand*\mciteBstWouldAddEndPuncttrue
  {\def\EndOfBibitem{\unskip.}}
\providecommand*\mciteBstWouldAddEndPunctfalse
  {\let\EndOfBibitem\relax}
\providecommand*\mciteSetBstMidEndSepPunct[3]{}
\providecommand*\mciteSetBstSublistLabelBeginEnd[3]{}
\providecommand*\EndOfBibitem{}
\mciteSetBstSublistMode{f}
\mciteSetBstMaxWidthForm{subitem}{(\alph{mcitesubitemcount})}
\mciteSetBstSublistLabelBeginEnd
  {\mcitemaxwidthsubitemform\space}
  {\relax}
  {\relax}

\bibitem[Kelly \latin{et~al.}(2007)Kelly, Baret, Taly, and
  Griffiths]{kelly2007miniaturizing}
Kelly,~B.~T.; Baret,~J.-C.; Taly,~V.; Griffiths,~A.~D. Miniaturizing Chemistry
  and Biology in Microdroplets. \emph{Chem. Commun.} \textbf{2007},
  1773--1788\relax
\mciteBstWouldAddEndPuncttrue
\mciteSetBstMidEndSepPunct{\mcitedefaultmidpunct}
{\mcitedefaultendpunct}{\mcitedefaultseppunct}\relax
\EndOfBibitem
\bibitem[Wang \latin{et~al.}(2009)Wang, Yang, and Li]{wang2009efficient}
Wang,~W.; Yang,~C.; Li,~C.~M. Efficient On-Demand Compound Droplet Formation:
  From Microfluidics to Microdroplets as Miniaturized Laboratories.
  \emph{Small} \textbf{2009}, \emph{5}, 1149--1152\relax
\mciteBstWouldAddEndPuncttrue
\mciteSetBstMidEndSepPunct{\mcitedefaultmidpunct}
{\mcitedefaultendpunct}{\mcitedefaultseppunct}\relax
\EndOfBibitem
\bibitem[Zhang \latin{et~al.}(2017)Zhang, Ettelaie, Yan, Zhang, Cheng, Binks,
  and Yang]{zhang2017ionic}
Zhang,~M.; Ettelaie,~R.; Yan,~T.; Zhang,~S.; Cheng,~F.; Binks,~B.~P.; Yang,~H.
  Ionic Liquid Droplet Microreactor for Catalysis Reactions Not at Equilibrium.
  \emph{J. Am. Chem. Soc.} \textbf{2017}, \emph{139}, 17387--17396\relax
\mciteBstWouldAddEndPuncttrue
\mciteSetBstMidEndSepPunct{\mcitedefaultmidpunct}
{\mcitedefaultendpunct}{\mcitedefaultseppunct}\relax
\EndOfBibitem
\bibitem[Guardingo \latin{et~al.}(2016)Guardingo, Busqu{\'e}, and
  Ruiz-Molina]{guardingo2016reactions}
Guardingo,~M.; Busqu{\'e},~F.; Ruiz-Molina,~D. Reactions in Ultra-Small
  Droplets by Tip-Assisted Chemistry. \emph{Chem. Commun.} \textbf{2016},
  \emph{52}, 11617--11626\relax
\mciteBstWouldAddEndPuncttrue
\mciteSetBstMidEndSepPunct{\mcitedefaultmidpunct}
{\mcitedefaultendpunct}{\mcitedefaultseppunct}\relax
\EndOfBibitem
\bibitem[Zhu and Power(2008)Zhu, and Power]{zhu2008lab}
Zhu,~Y.; Power,~B.~E. Lab-On-A-Chip \textit{In Vitro} Compartmentalization
  Technologies for Protein Studies. In \emph{Protein--Protein Interaction};
  Springer: Berlin, Heidelberg, 2008; Vol. 110; pp 81--114\relax
\mciteBstWouldAddEndPuncttrue
\mciteSetBstMidEndSepPunct{\mcitedefaultmidpunct}
{\mcitedefaultendpunct}{\mcitedefaultseppunct}\relax
\EndOfBibitem
\bibitem[Feng \latin{et~al.}(2018)Feng, Ueda, and Levkin]{feng2018droplet}
Feng,~W.; Ueda,~E.; Levkin,~P.~A. Droplet Microarrays: From Surface Patterning
  to High-Throughput Applications. \emph{Adv. Mater.} \textbf{2018}, \emph{30},
  1706111\relax
\mciteBstWouldAddEndPuncttrue
\mciteSetBstMidEndSepPunct{\mcitedefaultmidpunct}
{\mcitedefaultendpunct}{\mcitedefaultseppunct}\relax
\EndOfBibitem
\bibitem[Dittrich and Manz(2006)Dittrich, and Manz]{dittrich2006lab}
Dittrich,~P.~S.; Manz,~A. Lab-On-A-Chip: Microfluidics in Drug Discovery.
  \emph{Nat. Rev. Drug Discovery} \textbf{2006}, \emph{5}, 210--218\relax
\mciteBstWouldAddEndPuncttrue
\mciteSetBstMidEndSepPunct{\mcitedefaultmidpunct}
{\mcitedefaultendpunct}{\mcitedefaultseppunct}\relax
\EndOfBibitem
\bibitem[Ueda \latin{et~al.}(2012)Ueda, Geyer, Nedashkivska, and
  Levkin]{ueda2012dropletmicroarray}
Ueda,~E.; Geyer,~F.~L.; Nedashkivska,~V.; Levkin,~P.~A. DropletMicroarray:
  Facile Formation of Arrays of Microdroplets and Hydrogel Micropads for Cell
  Screening Applications. \emph{Lab Chip} \textbf{2012}, \emph{12},
  5218--5224\relax
\mciteBstWouldAddEndPuncttrue
\mciteSetBstMidEndSepPunct{\mcitedefaultmidpunct}
{\mcitedefaultendpunct}{\mcitedefaultseppunct}\relax
\EndOfBibitem
\bibitem[Bain \latin{et~al.}(2016)Bain, Pulliam, Thery, and
  Cooks]{bain2016accelerated}
Bain,~R.~M.; Pulliam,~C.~J.; Thery,~F.; Cooks,~R.~G. Accelerated Chemical
  Reactions and Organic Synthesis in Leidenfrost Droplets. \emph{Angew. Chem.,
  Int. Ed.} \textbf{2016}, \emph{55}, 10478--10482\relax
\mciteBstWouldAddEndPuncttrue
\mciteSetBstMidEndSepPunct{\mcitedefaultmidpunct}
{\mcitedefaultendpunct}{\mcitedefaultseppunct}\relax
\EndOfBibitem
\bibitem[Yan \latin{et~al.}(2016)Yan, Bain, and Cooks]{yan2016organic}
Yan,~X.; Bain,~R.~M.; Cooks,~R.~G. Organic Reactions in Microdroplets: Reaction
  Acceleration Revealed by Mass Spectrometry. \emph{Angew. Chem., Int. Ed.}
  \textbf{2016}, \emph{55}, 12960--12972\relax
\mciteBstWouldAddEndPuncttrue
\mciteSetBstMidEndSepPunct{\mcitedefaultmidpunct}
{\mcitedefaultendpunct}{\mcitedefaultseppunct}\relax
\EndOfBibitem
\bibitem[Li \latin{et~al.}(2016)Li, Yan, and Cooks]{li2016role}
Li,~Y.; Yan,~X.; Cooks,~R.~G. The Role of the Interface in Thin Film and
  Droplet Accelerated Reactions Studied by Competitive Substituent Effects.
  \emph{Angew. Chem., Int. Ed.} \textbf{2016}, \emph{55}, 3433--3437\relax
\mciteBstWouldAddEndPuncttrue
\mciteSetBstMidEndSepPunct{\mcitedefaultmidpunct}
{\mcitedefaultendpunct}{\mcitedefaultseppunct}\relax
\EndOfBibitem
\bibitem[Banerjee \latin{et~al.}(2017)Banerjee, Gnanamani, Yan, and
  Zare]{banerjee2017can}
Banerjee,~S.; Gnanamani,~E.; Yan,~X.; Zare,~R.~N. Can All Bulk-Phase Reactions
  Be Accelerated in Microdroplets? \emph{Analyst} \textbf{2017}, \emph{142},
  1399--1402\relax
\mciteBstWouldAddEndPuncttrue
\mciteSetBstMidEndSepPunct{\mcitedefaultmidpunct}
{\mcitedefaultendpunct}{\mcitedefaultseppunct}\relax
\EndOfBibitem
\bibitem[Yan \latin{et~al.}(2017)Yan, Cheng, and Zare]{yan2017two}
Yan,~X.; Cheng,~H.; Zare,~R.~N. Two-Phase Reactions in Microdroplets without
  the Use of Phase-Transfer Catalysts. \emph{Angew. Chem., Int. Ed.}
  \textbf{2017}, \emph{56}, 3562--3565\relax
\mciteBstWouldAddEndPuncttrue
\mciteSetBstMidEndSepPunct{\mcitedefaultmidpunct}
{\mcitedefaultendpunct}{\mcitedefaultseppunct}\relax
\EndOfBibitem
\bibitem[Kuksenok(2014)]{kuksenok2014chemical}
Kuksenok,~O. Chemical Synthesis in Small Spaces. \emph{Physics} \textbf{2014},
  \emph{7}, 4\relax
\mciteBstWouldAddEndPuncttrue
\mciteSetBstMidEndSepPunct{\mcitedefaultmidpunct}
{\mcitedefaultendpunct}{\mcitedefaultseppunct}\relax
\EndOfBibitem
\bibitem[Girod \latin{et~al.}(2011)Girod, Moyano, Campbell, and
  Cooks]{girod2011accelerated}
Girod,~M.; Moyano,~E.; Campbell,~D.~I.; Cooks,~R.~G. Accelerated Bimolecular
  Reactions in Microdroplets Studied by Desorption Electrospray Ionization Mass
  Spectrometry. \emph{Chem. Sci.} \textbf{2011}, \emph{2}, 501--510\relax
\mciteBstWouldAddEndPuncttrue
\mciteSetBstMidEndSepPunct{\mcitedefaultmidpunct}
{\mcitedefaultendpunct}{\mcitedefaultseppunct}\relax
\EndOfBibitem
\bibitem[Wei \latin{et~al.}(2017)Wei, Wleklinski, Ferreira, and
  Cooks]{wei2017reaction}
Wei,~Z.; Wleklinski,~M.; Ferreira,~C.; Cooks,~R.~G. Reaction Acceleration in
  Thin Films with Continuous Product Deposition for Organic Synthesis.
  \emph{Angew. Chem., Int. Ed.} \textbf{2017}, \emph{129}, 9514--9518\relax
\mciteBstWouldAddEndPuncttrue
\mciteSetBstMidEndSepPunct{\mcitedefaultmidpunct}
{\mcitedefaultendpunct}{\mcitedefaultseppunct}\relax
\EndOfBibitem
\bibitem[Narayan \latin{et~al.}(2005)Narayan, Muldoon, Finn, Fokin, Kolb, and
  Sharpless]{narayan2005water}
Narayan,~S.; Muldoon,~J.; Finn,~M.; Fokin,~V.~V.; Kolb,~H.~C.; Sharpless,~K.~B.
  “On Water”: Unique Reactivity of Organic Compounds in Aqueous Suspension.
  \emph{Angew. Chem., Int. Ed.} \textbf{2005}, \emph{44}, 3275--3279\relax
\mciteBstWouldAddEndPuncttrue
\mciteSetBstMidEndSepPunct{\mcitedefaultmidpunct}
{\mcitedefaultendpunct}{\mcitedefaultseppunct}\relax
\EndOfBibitem
\bibitem[Bain \latin{et~al.}(2017)Bain, Sathyamoorthi, and
  Zare]{bain2017droplet}
Bain,~R.~M.; Sathyamoorthi,~S.; Zare,~R.~N. “On-Droplet” Chemistry: The
  Cycloaddition of Diethyl Azodicarboxylate and Quadricyclane. \emph{Angew.
  Chem., Int. Ed.} \textbf{2017}, \emph{129}, 15279--15283\relax
\mciteBstWouldAddEndPuncttrue
\mciteSetBstMidEndSepPunct{\mcitedefaultmidpunct}
{\mcitedefaultendpunct}{\mcitedefaultseppunct}\relax
\EndOfBibitem
\bibitem[Carroll and Hidrovo(2013)Carroll, and
  Hidrovo]{carroll2013experimental}
Carroll,~B.; Hidrovo,~C. Experimental Investigation of Inertial Mixing in
  Colliding Droplets. \emph{Heat Transfer Eng.} \textbf{2013}, \emph{34},
  120--130\relax
\mciteBstWouldAddEndPuncttrue
\mciteSetBstMidEndSepPunct{\mcitedefaultmidpunct}
{\mcitedefaultendpunct}{\mcitedefaultseppunct}\relax
\EndOfBibitem
\bibitem[Lee \latin{et~al.}(2015)Lee, Kim, Nam, and Zare]{lee2015microdroplet}
Lee,~J.~K.; Kim,~S.; Nam,~H.~G.; Zare,~R.~N. Microdroplet Fusion Mass
  Spectrometry for Fast Reaction Kinetics. \emph{Proc. Natl. Acad. Sci. U. S.
  A.} \textbf{2015}, \emph{112}, 3898--3903\relax
\mciteBstWouldAddEndPuncttrue
\mciteSetBstMidEndSepPunct{\mcitedefaultmidpunct}
{\mcitedefaultendpunct}{\mcitedefaultseppunct}\relax
\EndOfBibitem
\bibitem[Davis \latin{et~al.}(2017)Davis, Jacobs, Houle, and
  Wilson]{davis2017colliding}
Davis,~R.~D.; Jacobs,~M.~I.; Houle,~F.~A.; Wilson,~K.~R. Colliding-Droplet
  Microreactor: Rapid On-Demand Inertial Mixing and Metal-Catalyzed Aqueous
  Phase Oxidation Processes. \emph{Anal. Chem.} \textbf{2017}, \emph{89},
  12494--12501\relax
\mciteBstWouldAddEndPuncttrue
\mciteSetBstMidEndSepPunct{\mcitedefaultmidpunct}
{\mcitedefaultendpunct}{\mcitedefaultseppunct}\relax
\EndOfBibitem
\bibitem[Nakatani \latin{et~al.}(1995)Nakatani, Chikama, Kim, and
  Kitamura]{nakatani1995droplet}
Nakatani,~K.; Chikama,~K.; Kim,~H.-B.; Kitamura,~N. Droplet-Size Dependence of
  the Electron Transfer Rate across the Single-Microdroplet/Water Interface.
  \emph{Chem. Phys. Lett.} \textbf{1995}, \emph{237}, 133--136\relax
\mciteBstWouldAddEndPuncttrue
\mciteSetBstMidEndSepPunct{\mcitedefaultmidpunct}
{\mcitedefaultendpunct}{\mcitedefaultseppunct}\relax
\EndOfBibitem
\bibitem[Nakatani \latin{et~al.}(1995)Nakatani, Suto, Wakabayashi, Kim, and
  Kitamura]{nakatani1995direct}
Nakatani,~K.; Suto,~T.; Wakabayashi,~M.; Kim,~H.-B.; Kitamura,~N. Direct
  Analyses of an Electrochemically Induced Dye Formation Reaction across a
  Single-Microdroplet/Water Interface. \emph{J. Phys. Chem.} \textbf{1995},
  \emph{99}, 4745--4749\relax
\mciteBstWouldAddEndPuncttrue
\mciteSetBstMidEndSepPunct{\mcitedefaultmidpunct}
{\mcitedefaultendpunct}{\mcitedefaultseppunct}\relax
\EndOfBibitem
\bibitem[Nakatani \latin{et~al.}(1996)Nakatani, Wakabayashi, Chikama, and
  Kitamura]{nakatani1996electrochemical}
Nakatani,~K.; Wakabayashi,~M.; Chikama,~K.; Kitamura,~N. Electrochemical
  Studies on Mass Transfer of Ferrocene Derivatives across a
  Single-Nitrobenzene-Microdroplet/Water Interface. \emph{J. Phys. Chem.}
  \textbf{1996}, \emph{100}, 6749--6754\relax
\mciteBstWouldAddEndPuncttrue
\mciteSetBstMidEndSepPunct{\mcitedefaultmidpunct}
{\mcitedefaultendpunct}{\mcitedefaultseppunct}\relax
\EndOfBibitem
\bibitem[Fallah-Araghi \latin{et~al.}(2014)Fallah-Araghi, Meguellati, Baret,
  El~Harrak, Mangeat, Karplus, Ladame, Marques, and
  Griffiths]{fallah2014enhanced}
Fallah-Araghi,~A.; Meguellati,~K.; Baret,~J.-C.; El~Harrak,~A.; Mangeat,~T.;
  Karplus,~M.; Ladame,~S.; Marques,~C.~M.; Griffiths,~A.~D. Enhanced Chemical
  Synthesis at Soft Interfaces: A Universal Reaction-Adsorption Mechanism in
  Microcompartments. \emph{Phys. Rev. Lett.} \textbf{2014}, \emph{112},
  028301\relax
\mciteBstWouldAddEndPuncttrue
\mciteSetBstMidEndSepPunct{\mcitedefaultmidpunct}
{\mcitedefaultendpunct}{\mcitedefaultseppunct}\relax
\EndOfBibitem
\bibitem[Baffou \latin{et~al.}(2014)Baffou, Polleux, Rigneault, and
  Monneret]{baffou2014super}
Baffou,~G.; Polleux,~J.; Rigneault,~H.; Monneret,~S. Super-Heating and
  Micro-Bubble Generation around Plasmonic Nanoparticles under cw Illumination.
  \emph{J. Phys. Chem. C} \textbf{2014}, \emph{118}, 4890--4898\relax
\mciteBstWouldAddEndPuncttrue
\mciteSetBstMidEndSepPunct{\mcitedefaultmidpunct}
{\mcitedefaultendpunct}{\mcitedefaultseppunct}\relax
\EndOfBibitem
\bibitem[Nam \latin{et~al.}(2017)Nam, Lee, Nam, and Zare]{nam2017abiotic}
Nam,~I.; Lee,~J.~K.; Nam,~H.~G.; Zare,~R.~N. Abiotic Production of Sugar
  Phosphates and Uridine Ribonucleoside in Aqueous Microdroplets. \emph{Proc.
  Natl. Acad. Sci. U. S. A.} \textbf{2017}, \emph{114}, 12396--12400\relax
\mciteBstWouldAddEndPuncttrue
\mciteSetBstMidEndSepPunct{\mcitedefaultmidpunct}
{\mcitedefaultendpunct}{\mcitedefaultseppunct}\relax
\EndOfBibitem
\bibitem[Vaida(2017)]{vaida2017prebiotic}
Vaida,~V. Prebiotic Phosphorylation Enabled by Microdroplets. \emph{Proc. Natl.
  Acad. Sci. U. S. A.} \textbf{2017}, \emph{114}, 12359--12361\relax
\mciteBstWouldAddEndPuncttrue
\mciteSetBstMidEndSepPunct{\mcitedefaultmidpunct}
{\mcitedefaultendpunct}{\mcitedefaultseppunct}\relax
\EndOfBibitem
\bibitem[Urban(2014)]{urban2014compartmentalised}
Urban,~P.~L. Compartmentalised Chemistry: From Studies on the Origin of Life to
  Engineered Biochemical Systems. \emph{New J. Chem.} \textbf{2014}, \emph{38},
  5135--5141\relax
\mciteBstWouldAddEndPuncttrue
\mciteSetBstMidEndSepPunct{\mcitedefaultmidpunct}
{\mcitedefaultendpunct}{\mcitedefaultseppunct}\relax
\EndOfBibitem
\bibitem[Lee \latin{et~al.}(2018)Lee, Samanta, Nam, Nam, and
  Zare]{lee2018spontaneous_bio}
Lee,~J.~K.; Samanta,~D.; Nam,~I.; Nam,~H.~G.; Zare,~R.~N. Spontaneous Reduction
  of Biomolecules on the Surface of Water Droplets. \emph{Biophys. J.}
  \textbf{2018}, \emph{114}, 542a\relax
\mciteBstWouldAddEndPuncttrue
\mciteSetBstMidEndSepPunct{\mcitedefaultmidpunct}
{\mcitedefaultendpunct}{\mcitedefaultseppunct}\relax
\EndOfBibitem
\bibitem[Lee \latin{et~al.}(2018)Lee, Samanta, Nam, and
  Zare]{lee2018spontaneous}
Lee,~J.~K.; Samanta,~D.; Nam,~H.~G.; Zare,~R.~N. Spontaneous Formation of Gold
  Nanostructures in Aqueous Microdroplets. \emph{Nat. Commun.} \textbf{2018},
  \emph{9}, 1562\relax
\mciteBstWouldAddEndPuncttrue
\mciteSetBstMidEndSepPunct{\mcitedefaultmidpunct}
{\mcitedefaultendpunct}{\mcitedefaultseppunct}\relax
\EndOfBibitem
\bibitem[Banerjee and Zare(2015)Banerjee, and Zare]{banerjee2015syntheses}
Banerjee,~S.; Zare,~R.~N. Syntheses of Isoquinoline and Substituted Quinolines
  in Charged Microdroplets. \emph{Angew. Chem., Int. Ed.} \textbf{2015},
  \emph{127}, 15008--15012\relax
\mciteBstWouldAddEndPuncttrue
\mciteSetBstMidEndSepPunct{\mcitedefaultmidpunct}
{\mcitedefaultendpunct}{\mcitedefaultseppunct}\relax
\EndOfBibitem
\bibitem[Lee \latin{et~al.}(2015)Lee, Banerjee, Nam, and
  Zare]{lee2015acceleration}
Lee,~J.~K.; Banerjee,~S.; Nam,~H.~G.; Zare,~R.~N. Acceleration of Reaction in
  Charged Microdroplets. \emph{Q. Rev. Biophys.} \textbf{2015}, \emph{48},
  437--444\relax
\mciteBstWouldAddEndPuncttrue
\mciteSetBstMidEndSepPunct{\mcitedefaultmidpunct}
{\mcitedefaultendpunct}{\mcitedefaultseppunct}\relax
\EndOfBibitem
\bibitem[Ingram \latin{et~al.}(2016)Ingram, Boeser, and Zare]{ingram2016going}
Ingram,~A.~J.; Boeser,~C.~L.; Zare,~R.~N. Going Beyond Electrospray: Mass
  Spectrometric Studies of Chemical Reactions in and on Liquids. \emph{Chem.
  Sci.} \textbf{2016}, \emph{7}, 39--55\relax
\mciteBstWouldAddEndPuncttrue
\mciteSetBstMidEndSepPunct{\mcitedefaultmidpunct}
{\mcitedefaultendpunct}{\mcitedefaultseppunct}\relax
\EndOfBibitem
\bibitem[Chen \latin{et~al.}(2016)Chen, Wan, and Badu-Tawiah]{chen2016picomole}
Chen,~S.; Wan,~Q.; Badu-Tawiah,~A.~K. Picomole-Scale Real-Time Photoreaction
  Screening: Discovery of the Visible-Light-Promoted Dehydrogenation of
  Tetrahydroquinolines under Ambient Conditions. \emph{Angew. Chem., Int. Ed.}
  \textbf{2016}, \emph{128}, 9491--9495\relax
\mciteBstWouldAddEndPuncttrue
\mciteSetBstMidEndSepPunct{\mcitedefaultmidpunct}
{\mcitedefaultendpunct}{\mcitedefaultseppunct}\relax
\EndOfBibitem
\bibitem[Jacobs \latin{et~al.}(2017)Jacobs, Davies, Lee, Davis, Houle, and
  Wilson]{jacobs2017exploring}
Jacobs,~M.~I.; Davies,~J.~F.; Lee,~L.; Davis,~R.~D.; Houle,~F.; Wilson,~K.~R.
  Exploring Chemistry in Microcompartments Using Guided Droplet Collisions in a
  Branched Quadrupole Trap Coupled to a Single Droplet, Paper Spray Mass
  Spectrometer. \emph{Anal. Chem.} \textbf{2017}, \emph{89}, 12511--12519\relax
\mciteBstWouldAddEndPuncttrue
\mciteSetBstMidEndSepPunct{\mcitedefaultmidpunct}
{\mcitedefaultendpunct}{\mcitedefaultseppunct}\relax
\EndOfBibitem
\bibitem[Gallo \latin{et~al.}(2019)Gallo, Farinha, Dinis, Emwas, Santana,
  Nielsen, Goddard, and Mishra]{gallo2019chemical}
Gallo,~A.; Farinha,~A.~S.; Dinis,~M.; Emwas,~A.-H.; Santana,~A.;
  Nielsen,~R.~J.; Goddard,~W.~A.; Mishra,~H. The Chemical Reactions in
  Electrosprays of Water Do Not Always Correspond to Those at the Pristine
  Air--Water Interface. \emph{Chem. Sci.} \textbf{2019}, \emph{10},
  2566--2577\relax
\mciteBstWouldAddEndPuncttrue
\mciteSetBstMidEndSepPunct{\mcitedefaultmidpunct}
{\mcitedefaultendpunct}{\mcitedefaultseppunct}\relax
\EndOfBibitem
\bibitem[Jacobs \latin{et~al.}(2018)Jacobs, Davis, Rapf, and
  Wilson]{jacobs2018studying}
Jacobs,~M.~I.; Davis,~R.~D.; Rapf,~R.~J.; Wilson,~K.~R. Studying Chemistry in
  Micro-Compartments by Separating Droplet Generation from Ionization. \emph{J.
  Am. Soc. Mass Spectrom.} \textbf{2018}, \emph{30}, 339--343\relax
\mciteBstWouldAddEndPuncttrue
\mciteSetBstMidEndSepPunct{\mcitedefaultmidpunct}
{\mcitedefaultendpunct}{\mcitedefaultseppunct}\relax
\EndOfBibitem
\bibitem[Zhang \latin{et~al.}(2015)Zhang, Lu, Tan, Bao, He, Sun, and
  Lohse]{Zhang2015}
Zhang,~X.; Lu,~Z.; Tan,~H.; Bao,~L.; He,~Y.; Sun,~C.; Lohse,~D. Formation of
  Surface Nanodroplets under Controlled Flow Conditions. \emph{Proc. Natl.
  Acad. Sci. U. S. A.} \textbf{2015}, \emph{112}, 9253--9257\relax
\mciteBstWouldAddEndPuncttrue
\mciteSetBstMidEndSepPunct{\mcitedefaultmidpunct}
{\mcitedefaultendpunct}{\mcitedefaultseppunct}\relax
\EndOfBibitem
\bibitem[Dyett \latin{et~al.}(2017)Dyett, Yu, and Zhang]{dyett2017formation}
Dyett,~B.; Yu,~H.; Zhang,~X. Formation of Surface Nanodroplets of Viscous
  Liquids by Solvent Exchange. \emph{Eur. Phys. J. E: Soft Matter Biol. Phys.}
  \textbf{2017}, \emph{40}, 1--6\relax
\mciteBstWouldAddEndPuncttrue
\mciteSetBstMidEndSepPunct{\mcitedefaultmidpunct}
{\mcitedefaultendpunct}{\mcitedefaultseppunct}\relax
\EndOfBibitem
\bibitem[Bao \latin{et~al.}(2015)Bao, Rezk, Yeo, and Zhang]{Bao2015}
Bao,~L.; Rezk,~A.~R.; Yeo,~L.~Y.; Zhang,~X. Highly Ordered Arrays of Femtoliter
  Surface Droplets. \emph{Small} \textbf{2015}, \emph{11}, 4850--4855\relax
\mciteBstWouldAddEndPuncttrue
\mciteSetBstMidEndSepPunct{\mcitedefaultmidpunct}
{\mcitedefaultendpunct}{\mcitedefaultseppunct}\relax
\EndOfBibitem
\bibitem[Lu \latin{et~al.}(2016)Lu, Peng, and Zhang]{Lu2016}
Lu,~Z.; Peng,~S.; Zhang,~X. Influence of Solution Composition on the Formation
  of Surface Nanodroplets by Solvent Exchange. \emph{Langmuir} \textbf{2016},
  \emph{32}, 1700--1706\relax
\mciteBstWouldAddEndPuncttrue
\mciteSetBstMidEndSepPunct{\mcitedefaultmidpunct}
{\mcitedefaultendpunct}{\mcitedefaultseppunct}\relax
\EndOfBibitem
\bibitem[Li \latin{et~al.}(2018)Li, Bao, Yu, and Zhang]{li2018formation}
Li,~M.; Bao,~L.; Yu,~H.; Zhang,~X. Formation of Multicomponent Surface
  Nanodroplets by Solvent Exchange. \emph{J. Phys. Chem. C} \textbf{2018},
  \emph{122}, 8647--8654\relax
\mciteBstWouldAddEndPuncttrue
\mciteSetBstMidEndSepPunct{\mcitedefaultmidpunct}
{\mcitedefaultendpunct}{\mcitedefaultseppunct}\relax
\EndOfBibitem
\bibitem[Hain \latin{et~al.}(2019)Hain, Handschuh-Wang, Wesner, Druzhinin, and
  Sch{\"o}nherr]{hain2019multimodal}
Hain,~N.; Handschuh-Wang,~S.; Wesner,~D.; Druzhinin,~S.~I.; Sch{\"o}nherr,~H.
  Multimodal Microscopy-Based Identification of Surface Nanobubbles. \emph{J.
  Colloid Interface Sci.} \textbf{2019}, \emph{547}, 162--170\relax
\mciteBstWouldAddEndPuncttrue
\mciteSetBstMidEndSepPunct{\mcitedefaultmidpunct}
{\mcitedefaultendpunct}{\mcitedefaultseppunct}\relax
\EndOfBibitem
\bibitem[Vincent and Marmottant(2017)Vincent, and
  Marmottant]{vincent2017statics}
Vincent,~O.; Marmottant,~P. On the Statics and Dynamics of Fully Confined
  Bubbles. \emph{J. Fluid Mech.} \textbf{2017}, \emph{827}, 194--224\relax
\mciteBstWouldAddEndPuncttrue
\mciteSetBstMidEndSepPunct{\mcitedefaultmidpunct}
{\mcitedefaultendpunct}{\mcitedefaultseppunct}\relax
\EndOfBibitem
\bibitem[Dyett \latin{et~al.}(2019)Dyett, Li, Zhao, and Zhang]{dyett2019}
Dyett,~B.~P.; Li,~M.; Zhao,~H.; Zhang,~X. Plasmonic Nanobubbles in
  “Armored” Surface Nanodroplets. \emph{J. Phys. Chem. C} \textbf{2019},
  \emph{123}, 29866--29874\relax
\mciteBstWouldAddEndPuncttrue
\mciteSetBstMidEndSepPunct{\mcitedefaultmidpunct}
{\mcitedefaultendpunct}{\mcitedefaultseppunct}\relax
\EndOfBibitem
\bibitem[Favelukis(2004)]{favelukis2004dynamics}
Favelukis,~M. Dynamics of Foam Growth: Bubble Growth in a Limited Amount of
  Liquid. \emph{Polym. Eng. Sci.} \textbf{2004}, \emph{44}, 1900--1906\relax
\mciteBstWouldAddEndPuncttrue
\mciteSetBstMidEndSepPunct{\mcitedefaultmidpunct}
{\mcitedefaultendpunct}{\mcitedefaultseppunct}\relax
\EndOfBibitem
\bibitem[Xu \latin{et~al.}(2010)Xu, Zhao, and Li]{xu2010numerical}
Xu,~X.; Zhao,~G.; Li,~H. Numerical Simulation of Bubble Growth in a Limited
  Amount of Liquid. \emph{J. Appl. Polym. Sci.} \textbf{2010}, \emph{116},
  1264--1271\relax
\mciteBstWouldAddEndPuncttrue
\mciteSetBstMidEndSepPunct{\mcitedefaultmidpunct}
{\mcitedefaultendpunct}{\mcitedefaultseppunct}\relax
\EndOfBibitem
\bibitem[Arefmanesh and Advani(1991)Arefmanesh, and
  Advani]{arefmanesh1991diffusion}
Arefmanesh,~A.; Advani,~S. Diffusion-Induced Growth of a Gas Bubble in a
  Viscoelastic Fluid. \emph{Rheol. Acta} \textbf{1991}, \emph{30},
  274--283\relax
\mciteBstWouldAddEndPuncttrue
\mciteSetBstMidEndSepPunct{\mcitedefaultmidpunct}
{\mcitedefaultendpunct}{\mcitedefaultseppunct}\relax
\EndOfBibitem
\bibitem[Doinikov and Marmottant(2018)Doinikov, and
  Marmottant]{doinikov2018natural}
Doinikov,~A.~A.; Marmottant,~P. Natural Oscillations of a Gas Bubble in a
  Liquid-Filled Cavity Located in a Viscoelastic Medium. \emph{J. Sound Vib.}
  \textbf{2018}, \emph{420}, 61--72\relax
\mciteBstWouldAddEndPuncttrue
\mciteSetBstMidEndSepPunct{\mcitedefaultmidpunct}
{\mcitedefaultendpunct}{\mcitedefaultseppunct}\relax
\EndOfBibitem
\bibitem[Doinikov \latin{et~al.}(2018)Doinikov, Dollet, and
  Marmottant]{doinikov2018cavitation}
Doinikov,~A.~A.; Dollet,~B.; Marmottant,~P. Cavitation in a Liquid-Filled
  Cavity Surrounded by an Elastic Medium: Intercoupling of Cavitation Events in
  Neighboring Cavities. \emph{Phys. Rev. E} \textbf{2018}, \emph{98},
  013108\relax
\mciteBstWouldAddEndPuncttrue
\mciteSetBstMidEndSepPunct{\mcitedefaultmidpunct}
{\mcitedefaultendpunct}{\mcitedefaultseppunct}\relax
\EndOfBibitem
\bibitem[Doinikov \latin{et~al.}(2018)Doinikov, Dollet, and
  Marmottant]{doinikov2018model}
Doinikov,~A.~A.; Dollet,~B.; Marmottant,~P. Model for the Growth and the
  Oscillation of a Cavitation Bubble in a Spherical Liquid-Filled Cavity
  Enclosed in an Elastic Medium. \emph{Phys. Rev. E} \textbf{2018}, \emph{97},
  013108\relax
\mciteBstWouldAddEndPuncttrue
\mciteSetBstMidEndSepPunct{\mcitedefaultmidpunct}
{\mcitedefaultendpunct}{\mcitedefaultseppunct}\relax
\EndOfBibitem
\bibitem[Vincent \latin{et~al.}(2012)Vincent, Marmottant, Quinto-Su, and
  Ohl]{vincent2012birth}
Vincent,~O.; Marmottant,~P.; Quinto-Su,~P.~A.; Ohl,~C.-D. Birth and Growth of
  Cavitation Bubbles within Water under Tension Confined in a Simple Synthetic
  Tree. \emph{Phys. Rev. Lett.} \textbf{2012}, \emph{108}, 184502\relax
\mciteBstWouldAddEndPuncttrue
\mciteSetBstMidEndSepPunct{\mcitedefaultmidpunct}
{\mcitedefaultendpunct}{\mcitedefaultseppunct}\relax
\EndOfBibitem
\bibitem[Vincent \latin{et~al.}(2014)Vincent, Marmottant, Gonzalez-Avila, Ando,
  and Ohl]{vincent2014fast}
Vincent,~O.; Marmottant,~P.; Gonzalez-Avila,~S.~R.; Ando,~K.; Ohl,~C.-D. The
  Fast Dynamics of Cavitation Bubbles within Water Confined in Elastic Solids.
  \emph{Soft Matter} \textbf{2014}, \emph{10}, 1455--1461\relax
\mciteBstWouldAddEndPuncttrue
\mciteSetBstMidEndSepPunct{\mcitedefaultmidpunct}
{\mcitedefaultendpunct}{\mcitedefaultseppunct}\relax
\EndOfBibitem
\bibitem[Shang \latin{et~al.}(2016)Shang, Cheng, Wang, Yu, Zhao, Chen, and
  Gu]{shang2016osmotic}
Shang,~L.; Cheng,~Y.; Wang,~J.; Yu,~Y.; Zhao,~Y.; Chen,~Y.; Gu,~Z. Osmotic
  Pressure-Triggered Cavitation in Microcapsules. \emph{Lab Chip}
  \textbf{2016}, \emph{16}, 251--255\relax
\mciteBstWouldAddEndPuncttrue
\mciteSetBstMidEndSepPunct{\mcitedefaultmidpunct}
{\mcitedefaultendpunct}{\mcitedefaultseppunct}\relax
\EndOfBibitem
\bibitem[Toutov \latin{et~al.}(2016)Toutov, Betz, Haibach, Romine, and
  Grubbs]{toutov2016sodium}
Toutov,~A.~A.; Betz,~K.~N.; Haibach,~M.~C.; Romine,~A.~M.; Grubbs,~R.~H. Sodium
  Hydroxide Catalyzed Dehydrocoupling of Alcohols with Hydrosilanes. \emph{Org.
  Lett.} \textbf{2016}, \emph{18}, 5776--5779\relax
\mciteBstWouldAddEndPuncttrue
\mciteSetBstMidEndSepPunct{\mcitedefaultmidpunct}
{\mcitedefaultendpunct}{\mcitedefaultseppunct}\relax
\EndOfBibitem
\bibitem[Zhang and Ducker(2008)Zhang, and Ducker]{zhang2008interfacial}
Zhang,~X.~H.; Ducker,~W. Interfacial Oil Droplets. \emph{Langmuir}
  \textbf{2008}, \emph{24}, 110--115\relax
\mciteBstWouldAddEndPuncttrue
\mciteSetBstMidEndSepPunct{\mcitedefaultmidpunct}
{\mcitedefaultendpunct}{\mcitedefaultseppunct}\relax
\EndOfBibitem
\bibitem[Deegan \latin{et~al.}(2000)Deegan, Bakajin, Dupont, Huber, Nagel, and
  Witten]{deeganPRE}
Deegan,~R.~D.; Bakajin,~O.; Dupont,~T.~F.; Huber,~G.; Nagel,~S.~R.;
  Witten,~T.~A. Contact Line Deposits in an Evaporating Drop. \emph{Phys. Rev.
  E} \textbf{2000}, \emph{62}, 756--765\relax
\mciteBstWouldAddEndPuncttrue
\mciteSetBstMidEndSepPunct{\mcitedefaultmidpunct}
{\mcitedefaultendpunct}{\mcitedefaultseppunct}\relax
\EndOfBibitem
\bibitem[Zhang \latin{et~al.}(2014)Zhang, Belova, Wang, Dong, and
  Moehwald]{zhang2014controlled}
Zhang,~L.; Belova,~V.; Wang,~H.; Dong,~W.; Moehwald,~H. Controlled Cavitation
  at Nano/Microparticle Surfaces. \emph{Chem. Mater.} \textbf{2014}, \emph{26},
  2244--2248\relax
\mciteBstWouldAddEndPuncttrue
\mciteSetBstMidEndSepPunct{\mcitedefaultmidpunct}
{\mcitedefaultendpunct}{\mcitedefaultseppunct}\relax
\EndOfBibitem
\bibitem[Langmuir(1916)]{langmuir1916constitution}
Langmuir,~I. The Constitution and Fundamental Properties of Solids and Liquids.
  Part I. Solids. \emph{J. Am. Chem. Soc.} \textbf{1916}, \emph{38},
  2221--2295\relax
\mciteBstWouldAddEndPuncttrue
\mciteSetBstMidEndSepPunct{\mcitedefaultmidpunct}
{\mcitedefaultendpunct}{\mcitedefaultseppunct}\relax
\EndOfBibitem
\bibitem[Epstein and Plesset(1950)Epstein, and Plesset]{epstein1950stability}
Epstein,~P.~S.; Plesset,~M.~S. On the Stability of Gas Bubbles in Liquid-Gas
  Solutions. \emph{J. Chem. Phys.} \textbf{1950}, \emph{18}, 1505--1509\relax
\mciteBstWouldAddEndPuncttrue
\mciteSetBstMidEndSepPunct{\mcitedefaultmidpunct}
{\mcitedefaultendpunct}{\mcitedefaultseppunct}\relax
\EndOfBibitem
\bibitem[Lohse and Zhang(2015)Lohse, and Zhang]{ZhangLohse2015}
Lohse,~D.; Zhang,~X. Surface Nanobubbles and Nanodroplets. \emph{Rev. Mod.
  Phys.} \textbf{2015}, \emph{87}, 981\relax
\mciteBstWouldAddEndPuncttrue
\mciteSetBstMidEndSepPunct{\mcitedefaultmidpunct}
{\mcitedefaultendpunct}{\mcitedefaultseppunct}\relax
\EndOfBibitem
\bibitem[Berry \latin{et~al.}(2015)Berry, Neeson, Dagastine, Chan, and
  Tabor]{OpenDrop2015}
Berry,~J.; Neeson,~M.; Dagastine,~R.; Chan,~D.; Tabor,~R. Measurement of
  Surface and Interfacial Tension Using Pendant Drop Tensiometry. \emph{J.
  Colloid Interface Sci.} \textbf{2015}, \emph{454}, 226--237\relax
\mciteBstWouldAddEndPuncttrue
\mciteSetBstMidEndSepPunct{\mcitedefaultmidpunct}
{\mcitedefaultendpunct}{\mcitedefaultseppunct}\relax
\EndOfBibitem
\bibitem[Van~der Walt \latin{et~al.}(2014)Van~der Walt, Sch{\"o}nberger,
  Nunez-Iglesias, Boulogne, Warner, Yager, Gouillart, and Yu]{van2014scikit}
Van~der Walt,~S.; Sch{\"o}nberger,~J.~L.; Nunez-Iglesias,~J.; Boulogne,~F.;
  Warner,~J.~D.; Yager,~N.; Gouillart,~E.; Yu,~T. Scikit-Image: Image
  Processing in Python. \emph{PeerJ} \textbf{2014}, \emph{2}, e453\relax
\mciteBstWouldAddEndPuncttrue
\mciteSetBstMidEndSepPunct{\mcitedefaultmidpunct}
{\mcitedefaultendpunct}{\mcitedefaultseppunct}\relax
\EndOfBibitem
\bibitem[Allan \latin{et~al.}(Zenodo:
  http://doi.org/10.5281/zenodo.60550)Allan, Caswell, Keim, and Van
  Der~Wel]{allan2016trackpy}
Allan,~D.; Caswell,~T.; Keim,~N.; Van Der~Wel,~C. \textit{trackpy }. Trackpy
  v0. 3.2., Zenodo: http://doi.org/10.5281/zenodo.60550; 2016\relax
\mciteBstWouldAddEndPuncttrue
\mciteSetBstMidEndSepPunct{\mcitedefaultmidpunct}
{\mcitedefaultendpunct}{\mcitedefaultseppunct}\relax
\EndOfBibitem
\end{mcitethebibliography}
	
\end{document}